# Systematic study of fully heavy-flavored tetraquarks $Q_1Q_2\bar{Q}_1\bar{Q}_2$ ($Q_{1,2} \in \{b, c\}$): Mass spectra, threshold analysis, and confrontation with LHC data

A. A. Atangana Likéné,[1,2,*] F. Rothen,[1,†] D. Nga Ongodo,[2,‡] G. H. Ben-Bolie,[2,§] and T. Golling[1,∥]

[1]*Department of Particle Physics, Faculty of Science, University of Geneva, P.O. Box 1205, Geneva, Switzerland*

[2]*Laboratory of Nuclear, Atomic and Molecular Physics, Department of Physics, Faculty of Science, University of Yaounde I, P.O. Box 812, Yaounde, Cameroon*

Experimental searches for fully heavy flavored tetraquark states are actively pursued at the LHC. The LHCb, CMS, and ATLAS collaborations have investigated fully charmed $cc\bar{c}\bar{c}$, doubly charmed-bottom $bc\bar{b}\bar{c}$ and fully bottom $bb\bar{b}\bar{b}$ tetraquarks. Fully charmed states are probed as intermediate resonances in processes like $p + p \to J/\psi J/\psi$ and $p + p \to J/\psi\mu^+\mu^-$ at $\sqrt{s} = 7, 8$, and 13 TeV. Notably, narrow structures such as $X(6200)$, $X(6900)$, and $X(7300)$ observed in the di-$J/\psi$ mass spectrum have ignited theoretical interest. In this study, we use a nonrelativistic model to perform a detailed study of the mass spectra for the ground ($1S$) and excited ($1P$, $2S$, $1D$, $2P$, $3S$, and $4S$) states of fully heavy tetraquarks $Q_1Q_2\bar{Q}_1\bar{Q}_2$ ($Q_{1,2} \in \{b, c\}$) in the diquark-antiquark picture. Tetraquarks are modeled as color-singlet bound states build from an axial-vector diquark ($[Q_1Q_2]_{s=1}$) and axial-vector antidiquark ($[\bar{Q}_1\bar{Q}_2]_{s=1}$) in the $\bar{\mathbf{3}}$ and $\mathbf{3}$ color representations, respectively, bound together by color forces. With this setup, the Schrödinger equation is solved numerically for a modified version of the Cornell potential, using the three-point difference central method. The spin-dependent terms are dealt with nonperturbatively and are used to describe the splitting structure of the diquark. The tetraquark states are investigated for a wide range of $J^{PC}$ quantum numbers including $0^{++}$, $1^{+-}$, $2^{++}$ for $S$-waves, $0^{-+}$, $1^{-\pm}$, $2^{-\pm}$, $3^{--}$ for $P$-waves, and $0^{++}$, $1^{+\pm}$, $2^{+\pm}$, $3^{+\pm}$, $4^{++}$ for $D$-waves. Following the same line as many previous works, we attempt, through this study of tetraquark mass spectra, to provide their possible signatures in narrow structures recently discovered in the di-$J/\psi$ production spectrum by the LHCb, CMS, and ATLAS collaborations. Our results provide precise mass predictions, analyze stability against strong fall-apart decays, and are directly compared with recent LHC observations and other relevant theoretical models. Our findings supports the interpretation of these discovered exotic heavy resonances as the different excitations of the fully charmed tetraquark, offering a crucial guide for ongoing experimental identification.

## I. INTRODUCTION

The landscape of hadron physics has extended significantly beyond the traditional quark model, with a novel class of particles garnering considerable interest. While the quark model predicts several stable combinations of valence quarks and antiquarks, for many decades, only two categories were definitively observed: baryons, consisting of three quarks ($qqq$), and mesons, consisting of a quark-antiquark pair ($q\bar{q}$). Other possible configurations such as tetraquarks ($qq\bar{q}\bar{q}$), pentaquarks ($qqqq\bar{q}$), hexaquarks ($qqq\bar{q}\bar{q}\bar{q}$), glueballs ($gg$), and hybrids ($q\bar{q}g$) were historically classified as "exotics" [1,2]. These states earn this designation because their quantum numbers, masses, decay channels, and widths defy explanation within the conventional meson or baryon frameworks [3–7]. The unambiguous exotic nature of some candidates is further confirmed by their nonzero electric charge, which necessitates a valence quark content beyond simple $q\bar{q}$ or $qqq$ combinations [8,9]. For years, the existence of such states remained speculative due to a lack of convincing experimental evidence. A turning point came in 2003 with the Belle Collaboration's discovery of the $X(3872)$ [10], a charmoniumlike state with a mass of $M_{X(3872)} = 3871.65 \pm 0.06$ MeV characterized by an extremely narrow width of $\Gamma_{X(3872)} = 1.19 \pm 0.21$ MeV [11] and isospin-violating decay ratios, $\mathrm{Br}(X(3872) \to \omega J/\psi)/\mathrm{Br}(X(3872) \to \pi^+\pi^- J/\psi) = 1.1 \pm 0.4$ [10,12]. Its properties are inconsistent with a pure $c\bar{c}$ charmonium interpretation but are compatible with a tetraquark structure like $cu\bar{c}\bar{u}$. Subsequently, the first explicitly exotic charged state, the $Z_c^{\pm}(4430)$, was discovered by LHCb in 2014 [8]. Its charge mandates a minimal valence content of four quarks, such as $cu\bar{c}\bar{d}$ or $cd\bar{c}\bar{u}$. The experimental landscape has since expanded to include several dozen exotic candidates, with reliably confirmed states such as the fully charmed tetraquark $cc\bar{c}\bar{c}$-X(6900) observed by LHCb in 2020 [13] and later by CMS [14] and ATLAS [15] and the pentaquarks $P_c^+(4380)$ and $P_c^+(4450)$ (with quark content $uudc\bar{c}$) reported by LHCb in 2015 [16] (see the recent review in Ref. [17]).

*Contact author: Andre.Atangana@etu.unige.ch, aandreaime@yahoo.fr
†Contact author: franck.rothen@unige.ch
‡Contact author: ngadieudonne@yahoo.fr
§Contact author: gbenbolie@yahoo.fr
∥Contact author: tobias.golling@unige.ch

Despite these discoveries, a unified theoretical description of exotic hadrons remains elusive. In the absence of a direct first-principles calculation of hadron spectra from QCD, theorists must rely on model assumptions about the structure and interactions within these complex systems. Consequently, diverse theoretical approaches, which posit different internal compositions and employ various nonperturbative methods, have been developed. Their predictions exhibit varying degrees of agreement with experimental data. Among the plethora of exotic states, this research focuses specifically on fully heavy tetraquarks $Q_1Q_2\bar{Q}_1\bar{Q}_2$ $(Q_{1,2} \in \{b, c\})$, composed of two heavy quarks $Q_1Q_2$ and two heavy antiquarks $\bar{Q}_1\bar{Q}_2$. This choice is motivated by the significant simplification it brings: the absence of light quarks narrows down the number of applicable theoretical approaches and reduces computational complexity. While numerous theoretical calculations within different models exist, consensus is lacking on which predicted states possess sufficiently long lifetimes for experimental detection. Concurrently, experimental searches are actively pursued at the LHC by the LHCb [13,18], CMS [14,19,20], and ATLAS [15] collaborations. For fully charmed $cc\bar{c}\bar{c}$ states, predicted to lie in the mass range of 5.8–7.4 GeV, searches have been conducted in processes like $p + p \to J/\psi(1S)J/\psi(1S)$, $p + p \to J/\psi(1S)\psi(2S)$, and $p + p \to J/\psi\mu^+\mu^-$ at center-of-mass energies of $\sqrt{s} = 7, 8$, and 13 TeV. In 2020, LHCb announced the discovery of the narrow $X(6900)$ resonance in the di-$J/\psi$ spectrum [13], a candidate for an excited $cc\bar{c}\bar{c}$ state, alongside broader structures near 6.4 and 7.2 GeV. These findings received preliminary confirmation from CMS [14] and ATLAS [15] in 2022. In contrast, the search for fully bottom $bb\bar{b}\bar{b}$ tetraquarks, predicted in the mass range of 18.4–18.8 GeV, has not yet yielded definitive evidence. Searches by LHCb [18] [in the range 17.5–20.0 GeV for processes $p + p \to \Upsilon(1S)\Upsilon(1S)$ and $p + p \to \Upsilon\mu^+\mu^-$] and CMS [19] (17.5–19.5 GeV and a narrow resonance search in 16.5–27 GeV) have been inconclusive, though an intriguing resonance at 18.15 GeV was reported by the ANDY Collaboration in Cu + Au collisions at the Relativistic Heavy Ion Collider (RHIC) [21], interpreted as a potential $bb\bar{b}\bar{b}$ state decaying to two $\Upsilon(1S)$ states. Thus, substantial experimental effort is still required to confirm fully bottom tetraquarks.

The theoretical interpretation of exotic states often centers on two primary structural pictures: the molecular and the compact tetraquark models. Hadronic molecules are loosely bound systems of two color-singlet mesons, held together by pion or other light meson exchanges. This scenario gained traction as the masses of several $XYZ$ states were found very close to relevant meson-antimeson thresholds. For instance, if the $\chi_{c1}(3872)$ has a binding energy $E$ of less than 200 keV relative to the $D^0\bar{D}^{*0}$ threshold, its size, estimated via $R = 1/\sqrt{2\mu E}$ (where $\mu$ is the reduced mass of the two-hadron system), would exceed 10 fm [22,23], characterizing it as an extended object. Conversely, compact tetraquarks are envisioned as tightly bound states of colored diquark $(qq)$ and antidiquark $(\bar{q}\bar{q})$ pairs, confined by gluonic forces in a manner analogous to how colored $q\bar{q}$ pairs form color-neutral mesons [23]. In this picture, the diquark and antidiquark are themselves colored objects (e.g., in a color antitriplet $\bar{\mathbf{3}}$ and triplet $\mathbf{3}$, respectively) whose colorless combination forms a tetraquark. Such configurations are not forbidden by QCD and, in fact, open a new window onto a potentially vast spectrum of compact hadrons beyond conventional mesons and baryons.

A tetraquark is fundamentally a bound state of two quarks and two antiquarks. Quarks come in six flavors, which can be categorized by mass relative to the QCD confinement scale ($\Lambda_{\mathrm{QCD}} \sim 200\,\mathrm{MeV}$): light quarks $(u, d, s)$ and heavy quarks $(c, b, t)$. The current masses are $M_u = 2.16^{+0.49}_{-0.26}$ MeV, $M_d = 4.67^{+0.48}_{-0.17}$ MeV, $M_s = 93.4^{+8.6}_{-3.4}$ MeV, $M_c = 1.27 \pm 0.02$ GeV, $M_b = 4.18^{+0.03}_{-0.02}$ GeV, and $M_t = 172.69 \pm 0.30$ GeV [11]. Our focus is on fully heavy tetraquarks. However, the top quark $(t)$ decays via the weak interaction on a timescale too short to form hadronic bound states [24] and is therefore excluded. Given the many possible flavor combinations, our nonrelativistic model can compute both ground and excited states for $QQ\bar{Q}\bar{Q}$ systems. Experimentally, the most promising and accessible candidates are the symmetric compositions: fully charmed $T_{4c} = cc\bar{c}\bar{c}$, doubly charmed bottom $T_{2bc} = bc\bar{b}\bar{c}$, and fully bottom $T_{4b} = bb\bar{b}\bar{b}$ tetraquarks. The preference for these combinations stems from their production mechanism: tetraquark formation is favored when it requires the production of only two heavy quark-antiquark pairs ($2c\bar{c}$, $c\bar{c} + b\bar{b}$, or $2b\bar{b}$). Other asymmetric combinations would necessitate the production of at least three pairs, a significantly less probable event. On the other hand, the use of a nonrelativistic formalism is justified for heavy tetraquark and pentaquark systems because the large masses of heavy quarks ($c$ and $b$) imply small quark velocities ($\frac{v}{c} \ll 1$), which suppress relativistic effects. Consequently, a Schrödinger-based approach with an effective potential provides a reliable description of the mass spectra, with relativistic corrections treated as subleading contributions [25–27].

Motivated by the observed charm-sector candidates and the open possibility for fully bottomed states, this work employs a nonrelativistic diquark-antidiquark model to study fully heavy flavored tetraquarks $Q_1 Q_2 \bar{Q}_1 \bar{Q}_2$ (where $Q_{1,2} \in \{c, b\}$). The absence of light quarks in the $T_{4c}$, $T_{2bc}$, and $T_{4b}$ systems makes a meson-molecule interpretation unlikely, as binding via pion or light vector meson exchange would be inefficient. If they exist, these systems are bound directly by nonperturbative QCD forces, and studying their spectra is crucial for a deeper understanding of these interactions. Conversely, if they do not exist, we must understand why QCD forbids or disfavors such configurations. We model these tetraquarks as nonrelativistic two-body systems, comprising a diquark (labeled "**i**") and an antidiquark (labeled "**j**"), interacting via a modified Cornell potential known as the nonrelativistic screened potential [27–31]. The diquark and antidiquark are chosen to be in color antitriplet ($\bar{\mathbf{3}}$) and triplet ($\mathbf{3}$) representations, respectively. The rationale for this potential is its proven success in describing highly excited heavy quarkonia $Q_1 \bar{Q}_2$. Although originally formulated for quarkonium, the screening of the confining interaction arises from vacuum polarization and string-breaking mechanisms intrinsic to the QCD vacuum, independent of the specific heavy-source configuration [32,33]. In the diquark-antidiquark picture, the $[QQ]$ and $[\bar{Q}\bar{Q}]$ clusters form a color $\bar{\mathbf{3}}$-$\mathbf{3}$ pair, which supports a confining flux tube analogous to that in quarkonium. At large separations, this flux tube is equally susceptible to screening by dynamical light quark-antiquark pair creation, leading to a softening (screening) of the linear potential. This effect is particularly important for radially or orbitally excited fully heavy tetraquarks, where the system may probe distances where a purely linear potential would be unrealistic. Therefore, employing the screened Cornell potential provides a more realistic effective interaction for the diquark-antidiquark system, consistent with previous applications in heavy baryon and multiquark studies. Furthermore, such potential models are amenable to continuous improvement [34] and their parameters can be refined using the latest experimental data on the hadron spectrum. In the nonrelativistic quark model, the two-body dynamics are described by the time-independent Schrödinger equation, which is solved numerically via the three-point central difference scheme [27,35,36].

This paper is organized as follows. In Sec. II, we introduce the theoretical framework employed to compute the masses $M_{Q_1 Q_2 \bar{Q}_1 \bar{Q}_2}$ of fully heavy tetraquark states $Q_1 Q_2 \bar{Q}_1 \bar{Q}_2 \equiv T_{(4Q)}(nL)$, with $Q_{1,2} \in \{b, c\}$. Section III outlines the relevant quantum numbers used in the mass calculations. The numerical results and their analysis are presented in Sec. IV, including comparisons with the corresponding strong decay thresholds. Finally, Sec. V summarizes the main findings of this work and presents our concluding remarks.

## II. PROBLEM

In the present work, we employ a nonrelativistic quark model to investigate the mass spectra of fully heavy tetraquark states $Q_1 Q_2 \bar{Q}_1 \bar{Q}_2$ ($Q_{1,2} \in \{c, b\}$), treating them as bound systems composed of axial-vector diquarks $[Q_1 Q_2]$ and antidiquarks $[\bar{Q}_1 \bar{Q}_2]$. Our analysis proceeds in two main steps. First, we calculate the mass spectra of the heavy diquark (color antitriplet) and antidiquark (color triplet) configurations within the proposed framework. Second, we employ the axial-vector diquark as a constituent to determine the mass spectra of the fully heavy tetraquark states. In this step, the tetraquark is modeled as a two-body system formed by the axial-vector diquark and antidiquark, whose interaction gives rise to an overall color-singlet state. A schematic illustration of this procedure is shown in Fig. 1. The diagram illustrates the diquark-antidiquark formation mechanism for fully heavy tetraquarks. Two heavy quarks $Q$ (colored spheres) bind via gluon exchange (blue wavy line) into a color-antitriplet axial-vector diquark (orange ellipse). Correspondingly, two heavy antiquarks $\bar{Q}$ bind into a color-triplet antidiquark (cyan ellipse). Subsequently, the diquark and antidiquark interact through additional gluon exchange (purple wavy line) and merge into a color-singlet tetraquark (magenta ellipse). This schematic representation applies to the all-charm $cc\bar{c}\bar{c}$, all-bottom $bb\bar{b}\bar{b}$, and bottom-charm $bc\bar{b}\bar{c}$ configurations.

Since the kinetic energies of the constituents in heavy-quark bound systems are significantly smaller than their rest-mass energies, a nonrelativistic treatment with a static potential offers a reliable approximation. Accordingly, we adopt a nonrelativistic Hamiltonian for two-body problems, expressed as [37]

$$\hat{H} = \sum_{i=1}^{2} \left[ m_i - \left( \frac{\hbar}{\sqrt{2m_i}} \vec{\nabla} \right)^2 \right] + \sum_{i \leq j=1} V_{ij}(r_{ij}). \quad (1)$$

We consider a two-body system in which particle "i" interacts with particle "j" through a potential $V_{ij}(r_{ij})$, where $r_{ij}$ denotes the relative separation between the two particles. In two-body systems governed by a central potential, it is advantageous to work in the center-of-mass (CM) frame. In this frame, spherical coordinates allow a natural separation between the radial and angular components of the wave function, and the kinetic energy operator involves the reduced mass of the system, $\mu$. To treat the two-body dynamics within the nonrelativistic quark model, we solve the time-independent Schrödinger equation:

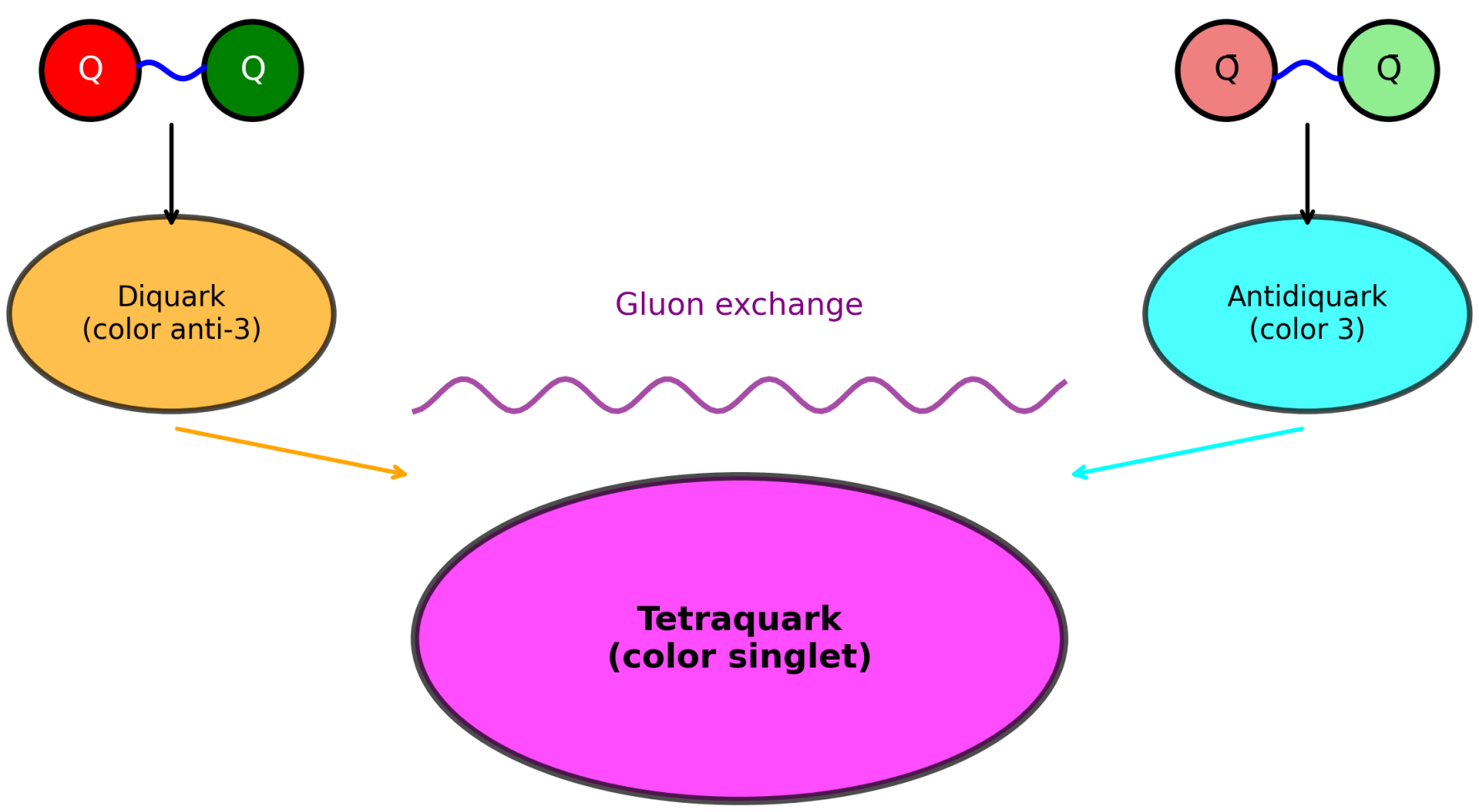

FIG. 1. Schematic illustration of the diquark-antidiquark formation mechanism for fully heavy tetraquarks $QQ\bar{Q}\,\bar{Q}$ ($Q \in \{c, b\}$).

$$\left[\frac{1}{2\mu}\left(-\frac{d^2}{dr_{ij}^2}+\frac{L(L+1)}{r_{ij}^2}\right)+V_{ij}(r_{ij})\right]\psi_{nL}(r_{ij})=E\psi_{nL}(r_{ij}),\quad \mu=\frac{m_i m_j}{m_i+m_j}. \tag{2}$$

One of the most widely used zeroth-order interquark potentials $V_{ij}^{(0)}(r_{ij})$ in heavy-quarkonium spectroscopy is the Cornell potential. It consists of a short-range Coulomb term $V_{ij}^{\mathrm{Coul}}(r_{ij})$ and a long-range linear confining term $V_{ij}^{\mathrm{Conf}}(r_{ij})$. The Coulomb contribution originates from one-gluon exchange with a Lorentz-vector structure, while the linear term accounts for confinement and is usually associated with a Lorentz-scalar structure. The Cornell potential is typically written as

$$V_{ij}^{(0)}(r_{ij}) = V_{ij}^{\mathrm{Coul}}(r_{ij}) + V_{ij}^{\mathrm{Conf}}(r_{ij}) = \kappa_s\frac{\alpha_s}{r_{ij}} + br_{ij}, \tag{3}$$

where $\kappa_s$ (the "color factor") encodes the color configuration of the two-body system and may take positive or negative values, $\alpha_s$ is the QCD strong coupling constant, and $b$ is the string tension that determines the strength of confinement. In the present work, instead of adopting the standard Cornell potential, we employ the nonrelativistic screened potential model [27–31] to compute the masses of fully heavy tetraquarks. In this approach, the usual linear confining term $br_{ij}$ is replaced by a screened interaction, and a constant term $\Lambda$ is added to represent a zero-point energy. With these modifications, the interquark potential takes the form

$$\begin{aligned} V_{ij}^{(0)}(r_{ij}) &= V_{ij}^{\mathrm{Vect}}(r_{ij}) + V_{ij}^{\mathrm{Scal}}(r_{ij}) \\ &= \kappa_s\frac{\alpha_s}{r_{ij}} + \frac{2b}{\eta}\exp\left(-\frac{\eta}{2}r_{ij}\right)\sinh\left(-\frac{\eta}{2}r_{ij}\right), \end{aligned} \tag{4}$$

$$V_{ij}^{\mathrm{Vect}}(r_{ij}) = V_{ij}^{\mathrm{Coul}}(r_{ij}) = \kappa_s\frac{\alpha_s}{r_{ij}}, \tag{5}$$

$$\begin{aligned} V_{ij}^{\mathrm{Scal}}(r_{ij}) &= V_{ij}^{\mathrm{Conf}}(r_{ij}) + \Lambda \\ &= \frac{2b}{\eta}\exp\left(-\frac{\eta}{2}r_{ij}\right)\sinh\left(-\frac{\eta}{2}r_{ij}\right). \end{aligned} \tag{6}$$

The reason behind this choice is that the linear potential, which is expected to be dominant at large distances, is screened or softened by the vacuum polarization effect of the dynamical light quark pairs [32,33]. Such a screening effect might be important for us to reasonably describe the higher radial and orbital excitations, which is one the specific objective of this work. Although originally formulated for quarkonium, the screening of the confining interaction arises from vacuum polarization and string-breaking mechanisms intrinsic to the QCD vacuum, independent of the specific heavy-source configuration [32,33]. In the diquark-antidiquark picture, the $[QQ]$ and $[\bar{Q}\bar{Q}]$ clusters form a color $\bar{\mathbf{3}}$-$\mathbf{3}$ pair, which supports a confining flux tube analogous to that in quarkonium. At large separations, this flux tube is equally susceptible to screening by dynamical light quark-antiquark pair creation, leading to a softening (screening) of the linear potential. This effect is particularly important for radially or orbitally excited fully heavy tetraquarks, where the system may probe distances where a purely linear potential would be unrealistic. Therefore, employing the screened Cornell potential provides a more realistic effective interaction for the diquark-antidiquark system.

In addition to the interquark potential $V_{ij}^{(0)}(r_{ij})$, the full interaction potential in Eq. (2) also contains a spin-dependent contribution, described by the Hamiltonian $\hat{H}_{ij}^{\mathrm{Spin}}$, such that

$$\begin{aligned} V_{ij}(r_{ij}) &= V_{ij}^{(0)}(r_{ij}) + \hat{H}_{ij}^{\mathrm{Spin}} \\ &= V_{ij}^{(0)}(r_{ij}) + \hat{H}_{ij}^{SS} + \hat{H}_{ij}^{LS} + \hat{H}_{ij}^{T}, \end{aligned} \tag{7}$$

where the spin-spin interaction $\hat{H}_{ij}^{SS}$, the spin-orbit interaction $\hat{H}_{ij}^{LS}$, and the tensor interaction $\hat{H}_{ij}^{T}$ are explicitly given by

$$\hat{H}_{ij}^{SS}(r_{ij}) = \Phi_{SS}(r_{ij})\vec{S}_i\cdot\vec{S}_j, \quad \text{with} \quad \vec{S}_i\cdot\vec{S}_j = \frac{1}{2}(\vec{S}^2 - \vec{S}_i^2 - \vec{S}_j^2), \tag{8}$$

$$\hat{H}_{ij}^{LS}(r_{ij}) = \Phi_{LS}(r_{ij})\vec{L}\cdot\vec{S}, \quad \text{with} \quad \vec{L}\cdot\vec{S} = \frac{1}{2}(\vec{J}^2 - \vec{L}^2 - \vec{S}^2), \tag{9}$$

$$\begin{aligned} \hat{H}_{ij}^{T}(r_{ij}) &= \Phi_T(r_{ij})\left[\frac{(\vec{S}_i\cdot\vec{r}_{ij})(\vec{S}_j\cdot\vec{r}_{ij})}{r_{ij}^2} - \frac{1}{3}(\vec{S}_i\cdot\vec{S}_j)\right] \\ &= \frac{1}{12}\Phi_T\hat{S}_{ij}, \end{aligned} \tag{10}$$

where the radial coefficients $\Phi_{SS}(r_{ij})$, $\Phi_{LS}(r_{ij})$, and $\Phi_T(r_{ij})$ come from the vector $V_{ij}^{\mathrm{Vect}}(r_{ij})$ and scalar $V_{ij}^{\mathrm{Scal}}(r_{ij})$ components of the interquark potential in Eqs. (4)–(6). These coefficients can be calculated as

$$\begin{aligned} \Phi_{SS}(r_{ij}) &= \frac{2}{3m_im_j}\vec{\nabla}V_{ij}^{\mathrm{Vect}}(r_{ij}) \\ &= \frac{1}{r_{ij}^2}\frac{\partial}{\partial r_{ij}}\left(r_{ij}^2\frac{\partial}{\partial r_{ij}}\right)\vec{\nabla}V_{ij}^{\mathrm{Vect}}(r_{ij}) \\ &= \begin{cases} 0, & \text{for } r_{ij} > 0 \\ -\frac{8\pi\alpha_s\kappa_s}{3m_im_j}\delta^{(3)}(r_{ij}), & \text{for } r_{ij}\to 0 \end{cases} \end{aligned} \tag{11}$$

$$\begin{aligned} \Phi_{LS}(r_{ij}) &= \frac{1}{2m_im_j}\frac{1}{r_{ij}}\left[3\frac{dV_{ij}^{\mathrm{Vect}}(r_{ij})}{dr_{ij}} - \frac{dV_{ij}^{\mathrm{Scal}}(r_{ij})}{dr_{ij}}\right] \\ &= -\frac{3\kappa_s\alpha_s}{2m_im_jr_{ij}^3} - \frac{b}{2m_im_jr_{ij}}\exp(-\eta r_{ij}), \end{aligned} \tag{12}$$

$$\begin{aligned} \Phi_T(r_{ij}) &= \frac{1}{12m_im_j}\left[\frac{1}{r_{ij}}\frac{dV_{ij}^{\mathrm{Vect}}(r_{ij})}{dr_{ij}} - \frac{d^2V_{ij}^{\mathrm{Vect}}(r_{ij})}{dr_{ij}^2}\right] \\ &= -\frac{\kappa_s\alpha_s}{4m_im_jr_{ij}^3}. \end{aligned} \tag{13}$$

The second term in the spin-orbit correction, proportional to the scalar contribution [see Eq. (13)], corresponds to a Thomas precession effect [38]. This arises from the assumption that the confining interaction originates from a Lorentz-scalar structure. It is worth noting that an additive constant term, denoted by Λ, is introduced in the interaction potential. Although this constant does not affect the radial coefficients since only derivatives of the potential enter their definitions, it remains an essential ingredient of the model. In practice, the inclusion of Λ results in a uniform shift of the entire spectrum and necessitates a readjustment of the model parameters in order to reproduce the heavy diquark masses, without leading to a genuine improvement in the quality of the fit. Nevertheless, such a constant term is commonly employed in phenomenological potential models to fix the absolute energy scale and to effectively absorb contributions from neglected short-distance dynamics and self-energy effects. This approach has been widely adopted in nonrelativistic quark models, notably in Cornell-type potentials used to describe heavy quarkonia [39,40], and is routinely applied when fitting meson spectra to extract parameters subsequently used in diquark and multi-quark studies [37,38,40–43]. The inclusion of Λ therefore ensures consistency with experimentally observed mass thresholds and maintains phenomenological continuity between meson, diquark, and tetraquark sectors.

The nonperturbative treatment of spin-dependent interactions constitutes a central feature of our approach. This choice significantly simplifies the implementation of the numerical method used in our computer code. Further technical details will be presented later. Moreover, following a framework widely adopted in quarkonium spectroscopy, improved agreement between predicted and observed $c\bar{c}$ states can be achieved by incorporating the spin-spin interaction directly into the zeroth-order potential [43–45]. This is implemented by replacing the Dirac delta function with a Gaussian smearing function, which introduces an additional parameter $\sigma$. With this prescription, the spin-spin interaction term takes the form

$$\hat{H}_{ij}^{SS} = -\frac{8\pi\kappa_s\alpha_s}{3m_im_j}\left(\frac{\sigma}{\sqrt{\pi}}\right)^3\exp(-\sigma^2r_{ij}^2)\vec{S}_i\cdot\vec{S}_j. \tag{14}$$

In the $|^{2S+1}L_J\rangle$ basis, the matrix elements for the spin-spin operator $\vec{S}_i\cdot\vec{S}_j$ is

$$\begin{aligned} \langle\vec{S}_i\cdot\vec{S}_j\rangle &= \frac{1}{2}\langle^{2S+1}L_J|\vec{S}^2 - \vec{S}_i^2 - \vec{S}_j^2|^{2S+1}L_J\rangle \\ &= \frac{1}{2}S(S+1) - \frac{3}{4}. \end{aligned} \tag{15}$$

The expectation value of the spin-orbit operator is

$$\langle \vec{L}\cdot\vec{S}\rangle = \frac{1}{2}\langle^{2S+1}L_J|\vec{J}^2-\vec{L}^2-\vec{S}^2|^{2S+1}L_J\rangle$$
$$= \frac{1}{2}[J(J+1)-L(L+1)-S(S+1)]. \quad (16)$$

The tensor interaction term requires additional algebraic manipulations. For convenience, we define the tensor operator with an additional factor of 12, which is absorbed by redefining its radial coefficient $\Phi_T(r_{ij})$ in Eq. (14) as follows:

$$\mathbf{S}_{12} \equiv 12\left(\frac{(\vec{S}_i\cdot\vec{r}_{ij})(\vec{S}_j\cdot\vec{r}_{ij})}{\vec{r}_{ij}^2} - \frac{1}{3}(\vec{S}_i\cdot\vec{S}_j)\right)$$
$$= 4\left[3\left(\vec{S}_i\cdot\frac{\vec{r}_{ij}}{r_{ij}}\right)\left(\vec{S}_j\cdot\frac{\vec{r}_{ij}}{r_{ij}}\right) - \vec{S}_i\cdot\vec{S}_j\right]. \quad (17)$$

The results for the diagonal matrix elements of the tensor operator between two spin-1/2 particles, as encountered in heavy quarkonium systems, are available in Refs. [46,47], and a more detailed discussion can be found in Ref. [48]. These diagonal matrix elements can be expressed as

$$\langle\hat{S}_{ij}\rangle = \frac{4\langle\vec{S}^2\vec{L}^2 - \frac{3}{2}\vec{L}\cdot\vec{S} - 3(\vec{L}\cdot\vec{S})^2\rangle}{(2L+3)(2L-1)}. \quad (18)$$

The expectation value of the tensor operator is nonzero only when $L\neq 0$ and $S=1$ (i.e., for triplet states) and for $J=L$, $J=L-1$, or $J=L+1$. By manipulating the spin operators and using the properties of spherical harmonics and Pauli matrices along with their respective eigenvalues, one obtains the following general expression. This result satisfies the above conditions and vanishes whenever $L=0$ or $S=0$, for any of the allowed values of $J$ and $L$:

$$\langle\mathbf{S}_{12}\rangle_{\frac{1}{2}\otimes\frac{1}{2}\to S=1,\ell\neq 0} = \begin{cases} -\frac{2\ell}{(2\ell+3)}, & \text{if } J=\ell+1,\\ +2, & \text{if } J=\ell,\\ -\frac{2(\ell+1)}{(2\ell-1)}, & \text{if } J=\ell-1. \end{cases} \quad (19)$$

The above result is not directly applicable to tetraquark states. In the tetraquark case, the diquark-antidiquark tensor interaction must be decomposed into a sum of four quark-antiquark tensor contributions, and the spin-1/2 tensor expression of Eq. (20) must be applied to each pair individually. We now explicitly act with the tensor operator on the angular part of the tetraquark wave function. We focus on the angular and spin degrees of freedom of the wave function and factorize the radial component from the angular one, which combines orbital and spin angular momenta coupled through Clebsch-Gordan coefficients. We label the two quarks inside the diquark [index ($i$)] by 1 and 2, and the two antiquarks inside the antidiquark [index ($j$)] by 3 and 4. Following the procedure presented in Ref. [38], to illustrate our treatment of the tensor interaction, we consider the specific example with total spin $S_T=2$, orbital angular momentum $L_T=1$, and total angular momentum $J_T=2$. The total spins of the diquark and antidiquark are denoted by $\vec{S}_i$ and $\vec{S}_j$, respectively. The spin states are written in the generic form $|S_T,M_{S_T}\rangle$, where $M_{S_T}$ is the magnetic projection. We present the wave functions both in the diquark-antidiquark spin basis and in the explicit four-particle spin basis, with the arrows always ordered as "1234." These constructions are inspired by Refs. [38,44,49–51] and references therein. Thus, for the quantum numbers specified above, the wave function can be written as

$$[(S_i=1)\otimes(S_j=1)\to(S_T=2)]\otimes(L_T=1)$$
$$\to |J_T,M_{J_T}\rangle = |2,2\rangle_{J_T}. \quad (20)$$

To construct the state $|J_T,M_{J_T}\rangle = |2,2\rangle_{J_T}$, we note that this configuration is characterized by $S_T=2$, $L_T=1$, $J_T=2$, and $M_{J_T}=2$. The first step consists of coupling the total spin $S_T$ with the orbital angular momentum $L_T$ to obtain the total angular momentum $J_T$. The corresponding Clebsch-Gordan decomposition is given by

$$|J_T=2,M_{J_T}=2\rangle = \sum_{M_{S_T}}\sum_{M_{L_T}}(\langle S_T=2,M_{S_T}|\otimes\langle L_T=1,M_{L_T}|$$
$$|J_T=2,M_{J_T}=2\rangle)|S_T=2,M_{S_T}\rangle\otimes|L_T=1,M_{L_T}\rangle. \quad (21)$$

The only possibilities for $M_{J_T}=M_{S_T}+M_{L_T}=2$ are $(M_{S_T},M_{L_T})=(2,0)$ and (1, 1). Thus,

$$|2,2\rangle_{J_T} = \sqrt{\frac{2}{3}}|S_T=2,M_{S_T}=2\rangle$$
$$\otimes|L_T=1,M_{L_T}=0\rangle - \sqrt{\frac{1}{3}}|S_T=2,M_{S_T}=1\rangle$$
$$\otimes|L_T=1,M_{L_T}=1\rangle. \quad (22)$$

Next, the state $|S_T,M_{S_T}\rangle$ must be written in terms of the diquark and antidiquark spin states. Restricting ourselves to axial-vector diquarks, we take $S_i=1$ and $S_j=1$. With the identification of quarks 1 and 2 as constituents of the diquark and quarks 3 and 4 as constituents of the anti-diquark, the allowed spin combinations are

$$|S_T = 2, M_{S_T} = 2\rangle = |S_i = 1, M_{S_i} = 1\rangle \otimes |S_j = 1, M_{S_j} = 1\rangle = |1,1\rangle_{12} \otimes |1,1\rangle_{34} \quad \text{for } M_{S_T} = 2, \tag{23}$$

$$\begin{aligned} |S_T = 2, M_{S_T} = 1\rangle &= \sqrt{\frac{1}{2}}[|S_i = 1, M_{S_i} = 1\rangle \otimes |S_j = 1, M_{S_j} = 0\rangle + |S_i = 1, M_{S_i} = 0\rangle \otimes |S_j = 1, M_{S_j} = 1\rangle] \\ &= \sqrt{\frac{1}{2}}(|1,1\rangle_{12} \otimes |1,0\rangle_{34} + |1,0\rangle_{12} \otimes |1,1\rangle_{34}) \quad \text{for } M_{S_T} = 1, \end{aligned} \tag{24}$$

and thus

$$|2,2\rangle_{J_T} = \sqrt{\frac{2}{3}}(|1,1\rangle_{12} \otimes |1,1\rangle_{34}) Y_1^0(\theta,\phi) - \sqrt{\frac{1}{3}}\left[\frac{|1,1\rangle_{12} \otimes |1,0\rangle_{34}}{\sqrt{2}} + \frac{|1,0\rangle_{12} \otimes |1,1\rangle_{34}}{\sqrt{2}}\right] Y_1^1(\theta,\phi). \tag{25}$$

Then, if we write spin states in quark basis,[1] we get

$$|1,1\rangle_{12} = |\uparrow\uparrow\rangle_{12}, \qquad |1,0\rangle_{12} = \frac{1}{\sqrt{2}}(|\uparrow\downarrow\rangle_{12} + |\downarrow\uparrow\rangle_{12}), \tag{26}$$

$$|1,1\rangle_{34} = |\uparrow\uparrow\rangle_{34}, \qquad |1,0\rangle_{34} = \frac{1}{\sqrt{2}}(|\uparrow\downarrow\rangle_{34} + |\downarrow\uparrow\rangle_{34}). \tag{27}$$

The arrows indicate the spins of the constituents in the fixed order (1, 2) and (3, 4), corresponding to the two quarks in the diquark and the two antiquarks in the antidiquark. An up (down) arrow represents the spin state $|\frac{1}{2},\frac{1}{2}\rangle$ $(|\frac{1}{2},-\frac{1}{2}\rangle)$. From this, Eq. (25) becomes

$$\begin{aligned} |2,2\rangle_{J_T} &= \sqrt{\frac{2}{3}}(|\uparrow\uparrow\rangle_{12} \otimes |\uparrow\uparrow\rangle_{34}) Y_1^0(\theta,\phi) - \sqrt{\frac{1}{6}}\left[|\uparrow\uparrow\rangle_{12} \otimes \left|\frac{\uparrow\downarrow + \downarrow\uparrow}{\sqrt{2}}\right\rangle_{34} + \left|\frac{\uparrow\downarrow + \downarrow\uparrow}{\sqrt{2}}\right\rangle_{12} \otimes |\uparrow\uparrow\rangle_{34}\right] Y_1^1(\theta,\phi) \\ &= \sqrt{\frac{2}{3}}(\uparrow\uparrow\uparrow\uparrow) Y_1^0(\theta,\phi) - \sqrt{\frac{1}{12}}(\uparrow\uparrow\uparrow\downarrow + \uparrow\uparrow\downarrow\uparrow + \uparrow\downarrow\uparrow\uparrow + \downarrow\uparrow\uparrow\uparrow) Y_1^1(\theta,\phi). \end{aligned} \tag{28}$$

We now explicitly apply the tensor operator to the angular wave function above. Within our approximations, this is equivalent to either applying the operator directly to the diquark-antidiquark pair (in the spin-1 basis) or considering the sum of the four tensor interactions between each quark-antiquark pair (in the spin-1/2 basis),[2] as expected from the rules of angular momentum algebra. Therefore, we have

$$\begin{aligned} \hat{S}_{ij} &= 12\left[\left(\vec{S}_i \cdot \frac{\vec{r}_{ij}}{r_{ij}}\right)\left(\vec{S}_j \cdot \frac{\vec{r}_{ij}}{r_{ij}}\right) - \frac{1}{3}(\vec{S}_i \cdot \vec{S}_j)\right] \\ &= \hat{S}_{14} + \hat{S}_{13} + \hat{S}_{24} + \hat{S}_{23}. \end{aligned} \tag{29}$$

To generalize the tensor interaction to the tetraquark case, we rewrite the tensor in a form that reproduces the familiar results for the specific case of two spin-$\frac{1}{2}$ particles, while also allowing a straightforward extension to other cases, such as the interaction between two spin-1 diquarks. The operator $\hat{S}_{12}$ in Eq. (20) is a rank-2 tensor, which can be expressed in terms of spin operators and spherical harmonics, as discussed in standard textbooks [52]. A detailed treatment of this method is provided in Ref. [48]. The functional form presented here does not rely on any specific relations or eigenvalues for spin-$\frac{1}{2}$ particles, but only on general properties from elementary angular momentum theory. One can express the unit vector $\hat{r}_{ij} = \frac{\vec{r}_{ij}}{r_{ij}}$ in spherical coordinates and the spin operators in Cartesian components, which can then be rearranged into raising, lowering, and $z$-component spin operators, together with spherical harmonics of $L = 2$, yielding:

[1]The "up" and "down" spin labels assigned to antiquarks refer to their spin states; however, for identical particles, we adopt the convention that ↑ corresponds to $S_z = +\frac{1}{2}$ and ↓ to $S_z = -\frac{1}{2}$, independently of whether the constituent is a quark or an antiquark, when working in the total spin-(z) basis.

[2]To illustrate this, we can write $S_i = S_1 + S_2$ and $S_j = S_3 + S_4$ and then expand the tensor between the diquark and antidiquark into a sum of four tensor operators acting between the individual quark-antiquark pairs.

$$\hat{S}_{ij} = \sqrt{\frac{256\pi}{5}} Y_2^0(\theta,\phi)\hat{S}_{1z}\hat{S}_{2z} - \sqrt{\frac{16\pi}{5}} Y_2^0(\theta,\phi)(\hat{S}_{1+}\hat{S}_{2-} + \hat{S}_{1-}\hat{S}_{2+}) + \sqrt{\frac{288\pi}{15}} Y_2^{-1}(\theta,\phi)(\hat{S}_{1z}\hat{S}_{2+} + \hat{S}_{1+}\hat{S}_{2z}) - \sqrt{\frac{288\pi}{15}} Y_2^1(\theta,\phi)(\hat{S}_{1z}\hat{S}_{2-} + \hat{S}_{1-}\hat{S}_{2z}) + \sqrt{\frac{288\pi}{15}} Y_2^{-2}(\theta,\phi)\hat{S}_{1+}\hat{S}_{2+} + \sqrt{\frac{288\pi}{15}} Y_2^2(\theta,\phi)\hat{S}_{1-}\hat{S}_{2-}. \quad (30)$$

From the expressions above, the expectation value of the tensor operator in the angular wave functions can be computed, as in Eq. (20), with the selection rules of the spherical harmonics used to determine the nonvanishing contributions.

## III. TETRAQUARK QUANTUM NUMBERS

In this section, we discuss the quantum numbers of tetraquarks, following Refs. [38,53,54]. The diquark-antidiquark basis can be used to label the possible tetraquark quantum numbers $J^{PC}$. We adopt the notation

$$|T_{4Q}\rangle = |S_i, S_j, S_T, L_T\rangle_{J_T}, \quad C_T = (-1)^{L_T+S_T}, \quad P_T = (-1)^{L_T}, \quad (31)$$

where $S_i$ and $S_j$ denote the total spins of the diquark and antidiquark, respectively; $S_T$ is the total spin of the tetraquark, arising from the coupling $S_i \otimes S_j$; $L_T$ is the orbital angular momentum between the diquark and antidiquark (within the two-body approximation); and $J_T$ is the total angular momentum of the tetraquark, obtained from the coupling $S_T \otimes L_T$. In Eq. (33), $C_T$ and $P_T$, respectively, stand for the charge-conjugation and parity quantum numbers of the tetraquark. Since we are interested in the T4c tetraquark, where the diquarks are composed of two charm quarks with spin 1 in the antitriplet color configuration, for the S-wave states, we have the following possibilities:

$$|0^{++}\rangle_{T4Q} = |S_{QQ}=1, S_{\bar{Q}\bar{Q}}=1, S_T=0, L_T=0\rangle_{J_T=0}$$
$$|1^{+-}\rangle_{T4Q} = |S_{QQ}=1, S_{\bar{Q}\bar{Q}}=1, S_T=1, L_T=0\rangle_{J_T=1}$$
$$|2^{++}\rangle_{T4Q} = |S_{QQ}=1, S_{\bar{Q}\bar{Q}}=1, S_T=2, L_T=0\rangle_{J_T=2}. \quad (32)$$

Note that all the $S$-wave tetraquark states discussed above possess positive parity. The introduction of the first orbital excitation produces a factor $(-1)$ in both the parity and the charge-conjugation quantum numbers. Consequently, all $P$-wave states (with $L_T = 1$) have negative parity and charge conjugation opposite to that of the $S$-wave states. For $P$- and $D$-wave tetraquark states within the diquark-antidiquark picture, we employ the same general notation and apply the standard parity and charge-conjugation relations. For $L_T = 1$ ($P$-waves), one can obtain $S_T = S_i \otimes S_j = 0, 1, 2$ from the coupling of two spin-1 diquarks. When $S_T = 0$, the total angular momentum is $J_T = 1$, since $S_T = 0$ and $L_T = 1$, leading to the quantum numbers $P_T = -1$, $C_T = -1$, and $J^{PC} = 1^{--}$. Therefore, the corresponding state is

$$|1^{--}\rangle_{T4Q} = |S_{QQ}=1, S_{\bar{Q}\bar{Q}}=1, S_T=0, L_T=1\rangle_{J_T=1}. \quad (33)$$

One can follow a similar procedure to find the quantum numbers in each of the other $P$ and $D$ waves. A summary of the possible quantum numbers for tetraquark states with $L_T = 0, 1, 3$ can be found in Table I. This table provides a complete mapping of quantum numbers for tetraquark states in the diquark-antidiquark model, from $S$-wave to $D$-wave excitations.

## IV. NUMERICAL RESULTS AND DISCUSSION

### A. Masses spectra of heavy diquark states

To study the mass spectra of diquark (a bound state of two quarks) or antidiquark (a bound state of two antiquarks), we derive the mass spectra of diquark (antidiquark) using the formula

TABLE I. Summary of possible quantum numbers and state notations for diquark-antidiquark tetraquark states.

| Waves | $L_T$ | $S_T$ | $J_T$ | $J^{PC}$ | State notation |
|---|---|---|---|---|---|
| S | 0 | 0 | 0 | $0^{++}$ | $\vert S_{QQ}=1, S_{\bar{Q}\bar{Q}}=1, S_T=0, L_T=0\rangle_{J_T=0}$ |
| | 0 | 1 | 1 | $1^{+-}$ | $\vert S_{QQ}=1, S_{\bar{Q}\bar{Q}}=1, S_T=1, L_T=0\rangle_{J_T=1}$ |
| | 0 | 2 | 2 | $2^{++}$ | $\vert S_{QQ}=1, S_{\bar{Q}\bar{Q}}=1, S_T=2, L_T=0\rangle_{J_T=2}$ |
| P | 1 | 0 | 1 | $1^{--}$ | $\vert S_{QQ}=1, S_{\bar{Q}\bar{Q}}=1, S_T=0, L_T=1\rangle_{J_T=1}$ |
| | 1 | 1 | 0 | $0^{-+}$ | $\vert S_{QQ}=1, S_{\bar{Q}\bar{Q}}=1, S_T=1, L_T=1\rangle_{J_T=0}$ |
| | 1 | 1 | 1 | $1^{-+}$ | $\vert S_{QQ}=1, S_{\bar{Q}\bar{Q}}=1, S_T=1, L_T=1\rangle_{J_T=1}$ |
| | 1 | 1 | 2 | $2^{-+}$ | $\vert S_{QQ}=1, S_{\bar{Q}\bar{Q}}=1, S_T=1, L_T=1\rangle_{J_T=2}$ |
| | 1 | 2 | 1 | $1^{--}$ | $\vert S_{QQ}=1, S_{\bar{Q}\bar{Q}}=1, S_T=2, L_T=1\rangle_{J_T=1}$ |
| | 1 | 2 | 2 | $2^{--}$ | $\vert S_{QQ}=1, S_{\bar{Q}\bar{Q}}=1, S_T=2, L_T=1\rangle_{J_T=2}$ |
| | 1 | 2 | 3 | $3^{--}$ | $\vert S_{QQ}=1, S_{\bar{Q}\bar{Q}}=1, S_T=2, L_T=1\rangle_{J_T=3}$ |
| D | 2 | 0 | 2 | $2^{++}$ | $\vert S_{QQ}=1, S_{\bar{Q}\bar{Q}}=1, S_T=0, L_T=2\rangle_{J_T=2}$ |
| | 2 | 1 | 1 | $1^{+-}$ | $\vert S_{QQ}=1, S_{\bar{Q}\bar{Q}}=1, S_T=1, L_T=2\rangle_{J_T=1}$ |
| | 2 | 1 | 2 | $2^{+-}$ | $\vert S_{QQ}=1, S_{\bar{Q}\bar{Q}}=1, S_T=1, L_T=2\rangle_{J_T=2}$ |
| | 2 | 1 | 3 | $3^{+-}$ | $\vert S_{QQ}=1, S_{\bar{Q}\bar{Q}}=1, S_T=1, L_T=2\rangle_{J_T=3}$ |
| | 2 | 2 | 0 | $0^{++}$ | $\vert S_{QQ}=1, S_{\bar{Q}\bar{Q}}=1, S_T=2, L_T=2\rangle_{J_T=0}$ |
| | 2 | 2 | 1 | $1^{++}$ | $\vert S_{QQ}=1, S_{\bar{Q}\bar{Q}}=1, S_T=2, L_T=2\rangle_{J_T=1}$ |
| | 2 | 2 | 2 | $2^{++}$ | $\vert S_{QQ}=1, S_{\bar{Q}\bar{Q}}=1, S_T=2, L_T=2\rangle_{J_T=2}$ |
| | 2 | 2 | 3 | $3^{++}$ | $\vert S_{QQ}=1, S_{\bar{Q}\bar{Q}}=1, S_T=2, L_T=2\rangle_{J_T=3}$ |
| | 2 | 2 | 4 | $4^{++}$ | $\vert S_{QQ}=1, S_{\bar{Q}\bar{Q}}=1, S_T=2, L_T=2\rangle_{J_T=4}$ |

$$M_{ij}^{\rm diquark} = m_i^{\rm quark} + m_j^{\rm quark} + E_{nLSJ}, \tag{34}$$

where $E_{nLSJ}$ is the binding energy of the quark "i" and the quark "j," which contains all the contributions of spin-dependent terms. This equation is similar to that of the quark-antiquark bound state system but with a different color factor, arising from the color antitriplet ($\bar{\mathbf{3}}$) representation for diquarks. On the basis of $SU(3)$ symmetry, when two quarks (or two antiquarks) are combined, we obtain $|QQ\rangle$: $\mathbf{3}\otimes\mathbf{3} = \bar{\mathbf{3}}\oplus\mathbf{6}$ in the fundamental $\mathbf{3}(\bar{\mathbf{3}})$ representation, which leads to a color factor $\kappa_s = -\frac{2}{3}$ [49,55–57]. It can be seen that the color factor $\kappa_s$ for the quark-antiquark bound state system in the singlet state and for the diquark (antidiquark) in the antitriplet (triplet) state is $-\frac{4}{3}$ and $-\frac{2}{3}$, respectively, which leads to the introduction of a factor of $\frac{1}{2}$. Since these factors arise from the color structure of the wave function, one would expect them to have an effect on the entire potential. So, to better describe the quark-quark interaction, we have extended this factor to the entire potential. Hence, the quark-quark (antiquark-antiquark) interaction potential was assumed to be $V_{QQ} = V_{\bar{Q}\bar{Q}} = \frac{1}{2}V_{Q\bar{Q}}$. Hence, we obtained the diquark masses by solving the Schrödinger equation with the interaction potential, which consists in a color Coulomb plus nonrelativistic screened potential plus a constant term.

In the beginning of our analysis, we first apply a $\chi$-squared fit of bottom- and charmed-meson masses to determine the free parameters of the potential. Through this approach, all free parameters are determined by minimizing $\chi^2$ given by [58,59]

$$\chi^2(N_p) = \sum_{i=1}^{n^{\rm data}} \frac{(M_i^{\rm data} - M_i^{\rm theory}(N_p))^2}{(\sigma_i^{\rm data})^2}. \tag{35}$$

In this equation, $M_i^{\rm data}$ stands for experimental estimate of diquark mass, and $M_i^{\rm theory}(N_p)$ refers to the respective theoretical prediction. Meanwhile, $\sigma_i^{\rm data}$ refers to the error of the corresponding experimental estimate. In Eq. (A1), $N_p$ indicates the total number of unknown parameters in the fit, and $n^{\rm data}$ refers to the number of data points. More details on this fitting procedure can be found in Refs. [60–62]. Considering the equation of diquark mass Eq. (36), the diquark masses were calculated considering the masses for diquark with various quark content, even those containing light quarks, just to ensure that the model can reproduce experimental data from different mass sector and is quite stable. Even though only the masses of the axial-vector heavy diquarks will be use to evaluate the tetraquark masses, we have also calculated the masses of scalar diquarks simply to ensure that the model can reproduce the masses of a wider range of diquarks. We have considered the available experimental data for the vector and pseudoscalar mesons as experimental estimates to detemine the masses of diquarks.

It is important to clarify the motivation behind comparing theoretically calculated diquark masses with experimentally measured meson masses, as presented in Table II. In our diquark-antidiquark framework, the diquark is treated as a colored constituent whose effective mass is derived by solving the two-body Schrödinger equation with a screened Cornell potential. The same potential model parameters (quark masses, coupling constant $\alpha_s$, string tension $b$, and screening parameter $\eta$) are calibrated by fitting to the experimentally known spectrum of selected mesons. Thus, the comparison with meson masses serves as an essential consistency check of our parameter set, ensuring that the model correctly reproduces established hadronic spectroscopy before being extended to exotic configurations. We emphasize that diquark masses are generally expected to be systematically higher than the masses of mesons with identical quark content. This is due to the weaker color attraction in the antitriplet ($\bar{\mathbf{3}}$) channel compared to the color-singlet channel of quark-antiquark systems, a feature consistently noted in previous diquark-based studies [63–65]. Nevertheless, the correspondence between diquark and meson masses remains phenomenologically informative because both systems are governed by the same nonperturbative QCD dynamics and share identical heavy quark constituents.

TABLE II. Theoretical and experimental estimates of the ground-state masses of axial-vector and scalar heavy diquarks. All masses are given in unit of MeV, and the mass of the nearest corresponding meson is taken as the experimental estimate for the heavy diquark masses.

| Heavy diquark | Nearest meson | Experimental values [11] estimate | Ref. [58] | This work | Relative error (%) |
|---|---|---|---|---|---|
| $[bu]_{s=0}$ | $B^-$ | $5279.34 \pm 0.12$ | 5269.21 | 5274.28 | 0.096 |
| $[bd]_{s=0}$ | $\bar{B}^0$ | $5279.65 \pm 0.12$ | 5269.21 | 5274.43 | 0.099 |
| $[bc]_{s=0}$ | $B_c^-$ | $6274.47 \pm 0.27$ | 6268.5 | 6271.70 | 0.044 |
| $[cu]_{s=0}$ | $D^0$ | $1864.84 \pm 0.05$ | 1859.48 | 1862.16 | 0.144 |
| $[cd]_{s=0}$ | $D$ | $1869.65 \pm 0.05$ | 1859.48 | 1864.57 | 0.272 |
| $[cs]_{s=0}$ | $D_s$ | $1968.35 \pm 0.07$ | 1972.29 | 1970.32 | 0.100 |
| $[bs]_{s=0}$ | $B_s$ | $5366.92 \pm 0.10$ | 5357.2 | 5362.04 | 0.091 |
| $[bs]_{s=1}$ | $B_s^*$ | $5415.8^{+1.5}_{-1.4}$ | 5399.3 | 5407.55 | 0.152 |
| $[cd]_{s=1}$ | $D^*$ | $2006.85 \pm 0.05$ | 2010.76 | 2008.81 | 0.098 |
| $[cs]_{s=1}$ | $D_s^*$ | $2114.2 \pm 0.4$ | 2120.9 | 2117.55 | 0.159 |
| $[cc]_{s=0}$ | $\eta_c$ | $2983.9 \pm 0.5$ | 2985.50 | 2984.70 | 0.027 |
| $[cc]_{s=1}$ | $J/\psi$ | $3096.90 \pm 0.006$ | 3089.78 | 3093.34 | 0.115 |
| $[bb]_{s=0}$ | $\eta_b$ | $9399.0 \pm 2.3$ | 9390.10 | 9394.40 | 0.049 |
| $[bb]_{s=1}$ | $\Upsilon$ | $9460.30 \pm 0.26$ | 9453.36 | 9456.83 | 0.037 |

The theoretical predictions for heavy diquark masses are presented in Table II and demonstrate remarkable quantitative agreement with experimental estimates derived from the nearest corresponding mesons. Several noteworthy features emerge from the analysis of our results. The calculated masses exhibit the expected ordering based on constituent quark masses. Bottom-containing diquarks (≈5.3–9.4 GeV) are significantly heavier than charm-containing systems (≈1.9–3.1 GeV), while mixed $[bc]$ diquarks occupy an intermediate mass range of ≈6.3 GeV. This mass hierarchy follows naturally from the heavier bottom quark mass relative to charm. A particularly noteworthy feature is the consistent spin-splitting between scalar ($S = 0$) and axial-vector ($S = 1$) diquark states. The computed mass differences, approximately 100 MeV for $[cc]$, 60 MeV for $[bb]$, and 45 MeV for $[bs]$ closely align with the experimental splittings observed in the corresponding meson pairs: ($\eta_c$, $J/\psi$), ($\eta_b$, $\Upsilon$), and ($B_s$, $B_s^*$), respectively. This agreement validates the nonperturbative treatment of spin-spin interactions within our theoretical framework. Additionally, our theoretical approach yields masses that systematically fall between the predictions of Amiri *et al.* [58] and the experimental meson values, achieving relative errors below 0.3% across all diquark configurations. The largest deviation occurs for the $[cd]_{s=0}$ diquark (0.272%), while the best agreement is observed for $[cc]_{s=0}$ (0.027%). The close correspondence between computed diquark masses and measured meson masses despite the expected upward shift due to color structure supports the reliability of our parameter set and provides a solid foundation for extending the model to fully heavy tetraquark systems.

Moreover, Figs. 2 and 3 present graphical comparisons of diquark masses across different spin states and illustrate the deviations between theoretical predictions and experimental estimates based on Particle Data Group (PDG) 2025 data [11]. These visualizations underscore the systematic accuracy of our approach and highlight the role of spin-dependent interactions in shaping the diquark spectrum. In summary, while diquarks are distinct colored objects expected to be heavier than their mesonic counterparts, the comparison with meson masses serves as a critical benchmark for validating our potential model parameters. The excellent agreement observed in spin splittings and mass trends reinforces the diquark-antidiquark picture as a consistent and predictive framework for investigating fully heavy tetraquark spectroscopy.

### B. Fully heavy $QQ\bar{Q}\bar{Q}$ ($Q \in \{c, b\}$) tetraquark masses

In this subsection, the tetraquark masses will be explicitly presented for the ground state and various radial and orbital excitations. The color factor should correspond to the color singlet; therefore, we will use $\kappa_s = -\frac{4}{3}$ and also the same parameters $\alpha_s$, $\eta$, $b$, and $\sigma$ obtained from the fit of the $[cc]$, $[bc]$, and $[bb]$ spectra. The fitted values of the model parameters are tabulated in Table III. The quark masses are $m_b = 5.138$ GeV, $m_c = 1.735$ GeV, $m_u \approx m_d = 0.302$ GeV, and $m_s = 0.455$ GeV. The calculation of the total mass of the fully heavy tetraquark systems will also be analogous to the diquark case:

$$M(T_{Q_1Q_2\bar{Q}_3\bar{Q}_4}) = M_{[Q_1Q_2]} + M_{[\bar{Q}_1\bar{Q}_2]} + E_{nLSJ}^{[Q_1Q_2]-[\bar{Q}_1\bar{Q}_2]}. \quad (36)$$

As previously mentioned, in this study, we consider only tetraquarks composed of axial-vector diquarks. Therefore, while $cc\bar{c}\bar{c}$ and $bb\bar{b}\bar{b}$ systems necessarily contain only axial-vector (anti)diquarks due to flavor symmetry, the $bc\bar{b}\bar{c}$ tetraquark, which could in principle accommodate scalar or mixed diquark configurations, is restricted here to the axial-vector/axial-vector case.

The calculated masses of the ground and excited states of fully heavy tetraquarks $T_{(4c)} = cc\bar{c}\bar{c}$, $T_{(2bc)} = bc\bar{b}\bar{c}$, and $T_{(4b)} = bb\bar{b}\bar{b}$, composed of axial-vector diquarks $[QQ]_{s=1}$, are presented in Table IV. These results are obtained within the framework of the Coulomb plus nonrelativistic screened potential model. The spectrum includes radial excitations up to $4S$ and orbital excitations up to $3P$ and $1D$, revealing a rich structure of possible tetraquark states. However, while the model predicts a large number of states, not all are likely to be observable experimentally due to rapid fall-apart decays into heavy meson pairs. The excitations considered here are restricted to those between the diquark and antidiquark, which are more favorable for experimental detection because they increase the spatial separation between heavy quarks and antiquarks, thereby suppressing decay probabilities. The mass spectra exhibit clear systematic trends across the three tetraquark systems. For the ground states ($1S$), the masses span from approximately 6.16 GeV for $cc\bar{c}\bar{c}$ to 12.81 GeV for $bc\bar{b}\bar{c}$ and 19.29 GeV for $bb\bar{b}\bar{b}$. These values reflect the heavy quark mass hierarchy and are consistent with expectations from heavy-quark symmetry. The spin splittings within each multiplet are relatively small, on the order of tens of MeV, indicating that spin-dependent interactions are suppressed compared to the central confining and Coulombic potentials, a characteristic feature of heavy-quark systems. As excitation energy increases, the level spacing decreases, particularly for higher radial and orbital excitations, illustrating the softening of the screened confining potential at larger diquark-antidiquark separations. The mass ranges for excited states extend up to about 7.33 GeV for $cc\bar{c}\bar{c}$, 13.27 GeV for $bc\bar{b}\bar{c}$, and 19.93 GeV for $bb\bar{b}\bar{b}$. These ranges encompass several recently observed exotic resonances, such as $X(6900)$, $X(7200)$, and $X(7300)$, suggesting that these states can be interpreted as excited tetraquark configurations. Notably, the model predicts a dense spectrum of high-spin states at higher excitations, such as the $1^5D_4$ state near 19.70 GeV for $bb\bar{b}\bar{b}$, which may lie below di-$\Upsilon$ thresholds and thus could be narrow due to suppressed decays.

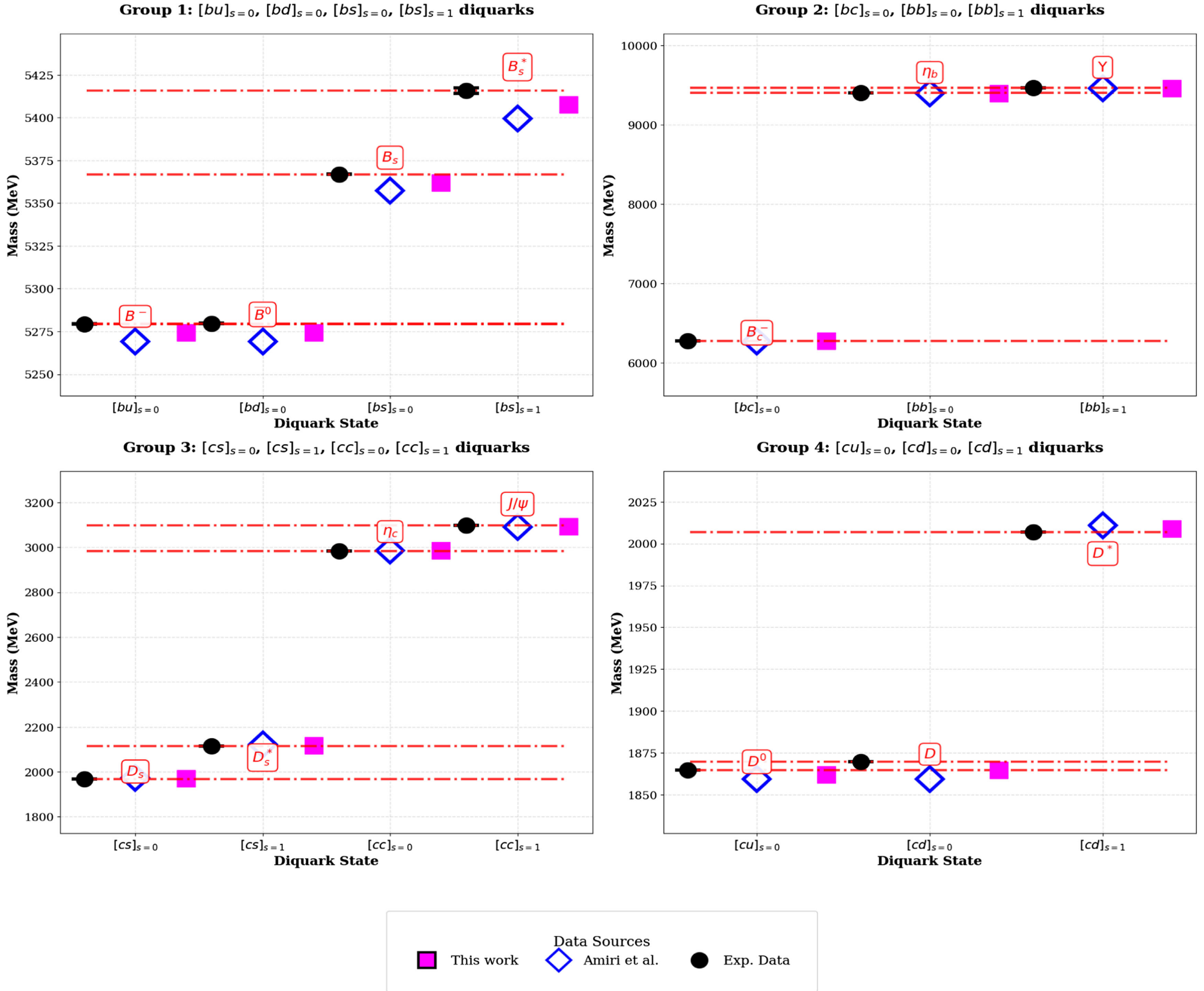

FIG. 2. Diquark masses comparison across different spin states.

The spectra also reveal systematic radial and orbital excitation patterns. Radial excitations ($2S$, $3S$, $4S$) exhibit a monotonic rise in mass with the principal quantum number $n$, characterized by excitation energies on the order of several hundred MeV. For instance, in $cc\bar{c}\bar{c}$, the $2S$ states are about 600 MeV above the $1S$ ground states, while $3S$ states are approximately 1.1 GeV higher. This pattern is analogous to conventional quarkonium spectra, though the absolute masses are significantly higher due to the four-quark composition. Orbital excitations ($1P$, $2P$, $3P$, and $1D$) introduce additional complexity, with multiplets split by tensor and spin-orbit interactions. The 1P states in $cc\bar{c}\bar{c}$, for example, lie around 6.60–6.64 GeV, overlapping with the region where several exotic charmoniumlike resonances have been reported. Additionally, an important feature is that orbital and radial excitations between the diquark and antidiquark can lead to narrower widths despite larger phase space for decay, due to centrifugal barriers or nodal structures in the wave functions that reduce overlap with fall-apart decay channels [66,67]. This suggests that some of the higher-mass states predicted here might be observable as relatively narrow resonances in experiments like Belle, LHCb, CMS, and ATLAS. However, alternative theoretical approaches such as hyperspherical harmonic expansions [68], string dynamics [69], and Hall-Post inequalities [70] have raised doubts about the stability of some fully heavy tetraquarks, indicating that not all predicted states may be bound. The calculated mass spectra offer a detailed mapping of potential fully heavy tetraquark states within the diquark-antidiquark model, providing key benchmarks for ongoing searches at collider experiments and for theoretical comparisons. However, the experimental detectability of these numerous predicted states hinges crucially on decay dynamics and their proximity to relevant meson-pair thresholds. This aspect is explored in the subsequent subsection.

FIG. 3. Deviation of theoretical predictions from experimental estimates.

TABLE III. Free parameters of the interquark potential for the fully heavy tetraquarks $Q_1Q_2\bar{Q}_1\bar{Q}_2$ ($Q_{1,2} \in \{c, b\}$). The quark masses are $m_c = 1310$ MeV and $m_b = 4660$ MeV.

| Tetraquark | Parameter | Value | Unit |
|---|---|---|---|
| $cc\bar{c}\bar{c}$ | $\alpha_s$ | 0.5209 | ... |
| | $b$ | 0.1866 | GeV$^2$ |
| | $\eta$ | 0.04879 | GeV |
| | $\sigma$ | 1.0967 | GeV |
| $bc\bar{b}\bar{c}$ | $\alpha_s$ | 0.3914 | ... |
| | $b$ | 0.1332 | GeV$^2$ |
| | $\eta$ | 0.03776 | GeV |
| | $\sigma$ | 1.5301 | GeV |
| $cc\bar{c}\bar{c}$ | $\alpha_s$ | 0.3411 | ... |
| | $b$ | 0.1109 | GeV$^2$ |
| | $\eta$ | 0.03142 | GeV |
| | $\sigma$ | 2.0146 | GeV |

## C. Threshold analysis and confrontation with LHC data

The decay dynamics and experimental signatures of fully heavy tetraquarks are critically determined by their masses relative to the relevant meson-meson dissociation thresholds. The phase space for decay, quantified by $\Delta = M_{QQ\bar{Q}\bar{Q}} - M_{\text{thr}}$, governs the dominant mechanism: a positive $\Delta$ allows a rapid, Okubo-Zweig-Iizuka rule-allowed "fall-apart" decay via quark rearrangement into two heavy quarkonia, typically producing a broad resonance; conversely, a state with $\Delta < 0$ for all possible two-meson channels is forced to decay via heavily suppressed processes,[3] resulting in a narrow width [65]. For the fully charmed $cc\bar{c}\bar{c}$ system, our detailed threshold analysis in Table V reveals a complex spectrum. The ground state $T_{(4c)0^{++}}(1S)$ with a mass of 6162 MeV lies intriguingly close to the di-$J/\psi$ threshold (6194 MeV), yielding $\Delta = -32$ MeV. This proximity suggests it could be a narrow state, making it a prime candidate for the $X(6200)$ resonance reported by ATLAS [15], a finding also supported by other analyses [71]. The first excited $S$-wave states, such as the $T_{(4c)2^{++}}(1S)$ around 6340 MeV, lie above the di-$J/\psi$ threshold ($\Delta \approx 150$ MeV) and could contribute to broad enhancements in the 6400 MeV region. More notably, the $1P$ multiplet including the $T_{(4c)0^{-+}}(1P)$ (6623 MeV), $T_{(4c)1^{-+}}(1P)$ (6629 MeV), and $T_{(4c)2^{-+}}(1P)$ (6638 MeV) features small positive $\Delta$ values (1–20 MeV) relative to channels like $J/\psi h_c$. This characteristic aligns with the broad $X(6600)$ structure observed by CMS and ATLAS [14,15], where our predicted masses of 6620–6640 MeV match the experimental mass of $6620 \pm 30^{+20}_{-10}$ MeV [15]. A significant concentration of predicted masses occurs near 6900 MeV, encompassing both $D$-wave and higher $S$-wave states. Specifically, the $T_{(4c)2^{++}}(1D)$ at 6887 MeV and several $1^5D_J$ states [e.g., $T_{(4c)3^{++}}(1D)$ at 6895 MeV] fall within the 50 MeV window of the well-established $X(6900)$ resonance [13–15]. This supports interpretations, such as those in Ref. [37], which assign the $X(6900)$ to a mixture of $2S$ and $1P$ states. Similarly, our predicted $3S$ states $T_{(4c)0^{++}}(3S)$ at 7189 MeV and $T_{(4c)2^{++}}(3S)$ at 7251 MeV coincide with the $X(7200)/X(7300)$ structures, corroborating the view that these are radial excitations. The consistently large positive $\Delta$ values ($\Delta > 500$ MeV) for these higher states explain their expected broad widths, as they can decay rapidly into various charmonium pairs.

[3]E.g., annihilation into gluons or radiative transitions.

TABLE IV. Calculated masses $M_{Q_1Q_2\bar{Q}_1\bar{Q}_2}$ ($Q_{1,2} \in \{b, c\}$) of the ground and excited states of $cc\bar{c}\,\bar{c}$, $bc\bar{b}\,\bar{c}$, and $bb\bar{b}\,\bar{b}$ tetraquarks composed from axial-vector diquarks $[cc]_{s=1}$, $[bc]_{s=1}$, and $[bb]_{s=1}$. The reported masses are given in unit of MeV.

| State | $n^{2S+1}L_J$ | $S$ | $L$ | $J$ | $J^{PC}$ | $M_{cc\bar{c}\,\bar{c}}$ | $M_{bc\bar{b}\,\bar{c}}$ | $M_{bb\bar{b}\,\bar{b}}$ |
|---|---|---|---|---|---|---|---|---|
| $T_{Q_1Q_2\bar{Q}_1\bar{Q}_2}(1S)$ | $1^1S_0$ | 0 | 0 | 0 | $0^{++}$ | 6162 | 12810 | 19286 |
| | $1^3S_1$ | 1 | 0 | 1 | $1^{+-}$ | 6243 | 12827 | 19292 |
| | $1^5S_2$ | 2 | 0 | 2 | $2^{++}$ | 6339 | 12855 | 19302 |
| $T_{Q_1Q_2\bar{Q}_1\bar{Q}_2}(1P)$ | $1^1P_1$ | 0 | 1 | 1 | $1^{--}$ | 6603 | 13075 | 19508 |
| | $1^3P_0$ | 1 | 1 | 0 | $0^{-+}$ | 6600 | 13072 | 19505 |
| | $1^3P_1$ | 1 | 1 | 1 | $1^{-+}$ | 6606 | 13075 | 19507 |
| | $1^3P_2$ | 1 | 1 | 2 | $2^{-+}$ | 6616 | 13080 | 19511 |
| | $1^5P_1$ | 2 | 1 | 1 | $1^{--}$ | 6607 | 13075 | 19506 |
| | $1^5P_2$ | 2 | 1 | 2 | $2^{--}$ | 6620 | 13081 | 19510 |
| | $1^5P_3$ | 2 | 1 | 3 | $3^{--}$ | 6636 | 13088 | 19517 |
| $T_{Q_1Q_2\bar{Q}_1\bar{Q}_2}(2S)$ | $2^1S_0$ | 0 | 0 | 0 | $0^{++}$ | 6754 | 13119 | 19652 |
| | $2^3S_1$ | 1 | 0 | 1 | $1^{+-}$ | 6788 | 13128 | 19654 |
| | $2^5S_2$ | 2 | 0 | 2 | $2^{++}$ | 6840 | 13144 | 19659 |
| $T_{Q_1Q_2\bar{Q}_1\bar{Q}_2}(1D)$ | $1^1D_2$ | 0 | 2 | 2 | $2^{++}$ | 6893 | 13178 | 19687 |
| | $1^3D_1$ | 1 | 2 | 1 | $1^{+-}$ | 6881 | 13171 | 19682 |
| | $1^3D_2$ | 1 | 2 | 2 | $2^{+-}$ | 6892 | 13176 | 19686 |
| | $1^3D_3$ | 1 | 2 | 3 | $3^{+-}$ | 6904 | 13183 | 19692 |
| | $1^5D_0$ | 2 | 2 | 0 | $0^{++}$ | 6871 | 13165 | 19677 |
| | $1^5D_1$ | 2 | 2 | 1 | $1^{++}$ | 6876 | 13168 | 19679 |
| | $1^5D_2$ | 2 | 2 | 2 | $2^{++}$ | 6887 | 13173 | 19683 |
| | $1^5D_3$ | 2 | 2 | 3 | $3^{++}$ | 6901 | 13180 | 19689 |
| | $1^5D_4$ | 2 | 2 | 4 | $4^{++}$ | 6917 | 13189 | 19696 |
| $T_{Q_1Q_2\bar{Q}_1\bar{Q}_2}(2P)$ | $2^1P_1$ | 0 | 1 | 1 | $1^{--}$ | 7063 | 13200 | 19792 |
| | $2^3P_0$ | 1 | 1 | 0 | $0^{-+}$ | 7072 | 13203 | 19793 |
| | $2^3P_1$ | 1 | 1 | 1 | $1^{-+}$ | 7071 | 13203 | 19793 |
| | $2^3P_2$ | 1 | 1 | 2 | $2^{-+}$ | 7070 | 13203 | 19794 |
| | $2^5P_1$ | 2 | 1 | 1 | $1^{--}$ | 7085 | 13206 | 19795 |
| | $2^5P_2$ | 2 | 1 | 2 | $2^{--}$ | 7085 | 13207 | 19795 |
| | $2^5P_3$ | 2 | 1 | 3 | $3^{--}$ | 7084 | 13208 | 19796 |
| $T_{Q_1Q_2\bar{Q}_1\bar{Q}_2}(3S)$ | $3^1S_0$ | 0 | 0 | 0 | $0^{++}$ | 7231 | 13230 | 19913 |
| | $3^3S_1$ | 1 | 0 | 1 | $1^{+-}$ | 7259 | 13238 | 19915 |
| | $3^5S_2$ | 2 | 0 | 2 | $2^{++}$ | 7305 | 13252 | 19919 |
| $T_{Q_1Q_2\bar{Q}_1\bar{Q}_2}(3P)$ | $3^1P_0$ | 0 | 1 | 0 | $0^{-+}$ | 7137 | 13220 | 19916 |
| | $3^3P_0$ | 1 | 1 | 0 | $0^{-+}$ | 7042 | 13215 | 19866 |
| | $3^3P_1$ | 1 | 1 | 1 | $1^{-+}$ | 7112 | 13218 | 19890 |
| | $3^3P_2$ | 1 | 1 | 2 | $2^{-+}$ | 7183 | 13222 | 19912 |
| | $3^5P_1$ | 2 | 1 | 1 | $1^{--}$ | 7201 | 13225 | 19973 |
| | $3^5P_2$ | 2 | 1 | 2 | $2^{--}$ | 7184 | 13226 | 19911 |
| | $3^5P_3$ | 2 | 1 | 3 | $3^{--}$ | 7139 | 13230 | 19944 |
| $T_{Q_1Q_2\bar{Q}_1\bar{Q}_2}(4S)$ | $4^1S_0$ | 0 | 0 | 0 | $0^{++}$ | 7203 | 13260 | 19281 |
| | $4^3S_1$ | 1 | 0 | 1 | $1^{+-}$ | 7214 | 13265 | 19284 |
| | $4^3S_2$ | 2 | 0 | 2 | $2^{++}$ | 7226 | 13270 | 19286 |

For the fully bottom $bb\bar{b}\bar{b}$ system, the scenario is markedly different. As shown in Table VI, all calculated masses lie significantly above the lowest di-$\Upsilon(1S)$ threshold (18921 MeV). The $\Delta$ values are substantial, ranging from ≈360 MeV for the $1S$ ground state to nearly 1 GeV for higher excitations, placing our predicted mass spectrum in the range of approximately 19.3 to 19.9 GeV. This unambiguously indicates that these tetraquarks are unbound and would decay rapidly via fall-apart processes, resulting in very broad resonances exceedingly difficult to detect in the clean $\mu^+\mu^-$ channel. This theoretical prediction is consistent with the null results from dedicated searches for the fully bottom tetraquark in the process $p + p \to T_{(4b)} \to \Upsilon(1S)\Upsilon(1S)$ in the mass range 17.5–20 GeV, which covers our predicted region. More details can be found in Refs. [18–20]. Moreover, lattice calculations [72] do not find fully bottom tetraquark bound states in this mass region. It also provides context for the nonobservation of the resonance at 18150 MeV reported by the ANDY Collaboration in heavy-ion collisions [21], which was tentatively interpreted as a $bb\bar{b}\bar{b}$ state; our higher mass predictions suggest such a state would be broad. However, a notable exception exists for high-spin, high-orbital-excitation states. Some of our $1D$ states including $T_{(4b)3^{++}}(19687)(1D)$ and $T_{(4b)4^{++}}(19694)(1D)$ lie very

TABLE V. Calculated masses of the ground and excited $T_{(4c)} = cc\bar{c}\,\bar{c}$ states composed from axial-vector diquarks and the corresponding meson-meson thresholds $M_{\rm thr}$ $\Delta = M_{cc\bar{c}\,\bar{c}} - M_{\rm thr}$. Masses are given in the unit of MeV.

| State | $n^{2S+1}L_J$ | $S$ | $L$ | $J^{PC}$ | $M_{cc\bar{c}\,\bar{c}}$ | $M_{\rm thr}$ | $\Delta$ | Meson pair |
|---|---|---|---|---|---|---|---|---|
| $T_{(4c)0^{++}}(6162)(1S)$ | $1^1S_0$ | 0 | 0 | $0^{++}$ | 6162 | 5968 | 194 | $\eta_c(1S)\eta_c(1S)$ |
| | | | | | | 6194 | −32 | $J/\psi(1S)J/\psi(1S)$ |
| $T_{(4c)1^{+-}}(6243)(1S)$ | $1^3S_1$ | 1 | 0 | $1^{+-}$ | 6243 | 6081 | 162 | $\eta_c(1S)J/\psi(1S)$ |
| $T_{(4c)2^{++}}(6343)(1S)$ | $1^5S_2$ | 2 | 0 | $2^{++}$ | 6343 | 6194 | 149 | $J/\psi(1S)J/\psi(1S)$ |
| | | | | | | 6509 | 108 | $\eta_c(1S)h_c(1P)$ |
| $T_{(4c)1^{--}}(6617)(1P)$ | $1^1P_1$ | 0 | 1 | $1^{--}$ | 6617 | 6512 | 105 | $J/\psi(1S)\chi_{c0}(1P)$ |
| | | | | | | 6608 | 9 | $J/\psi(1S)\chi_{c1}(1P)$ |
| $T_{(4c)0^{-+}}(6623)(1P)$ | $1^3P_0$ | 1 | 1 | $0^{-+}$ | 6623 | 6399 | 224 | $\eta_c(1S)\chi_{c0}(1P)$ |
| | | | | | | 6622 | 1 | $J/\psi(1S)h_c(1P)$ |
| $T_{(4c)1^{-+}}(6629)(1P)$ | $1^3P_1$ | 1 | 1 | $1^{-+}$ | 6629 | 6495 | 134 | $\eta_c(1S)\chi_{c1}(1P)$ |
| | | | | | | 6622 | 7 | $J/\psi(1S)h_c(1P)$ |
| $T_{(4c)2^{-+}}(6638)(1P)$ | $1^3P_2$ | 1 | 1 | $2^{-+}$ | 6638 | 6540 | 98 | $\eta_c(1S)\chi_{c2}(1P)$ |
| | | | | | | 6622 | 16 | $J/\psi(1S)h_c(1P)$ |
| | | | | | | 6509 | 121 | $\eta_c(1S)h_c(1P)$ |
| $T_{(4c)1^{--}}(6630)(1P)$ | $1^5P_1$ | 2 | 1 | $1^{--}$ | 6630 | 6512 | 118 | $J/\psi(1S)\chi_{c0}(1P)$ |
| | | | | | | 6608 | 22 | $J/\psi(1S)\chi_{c1}(1P)$ |
| | | | | | | 6653 | −23 | $J/\psi(1S)\chi_{c2}(1P)$ |
| $T_{(4c)2^{--}}(6642)(1P)$ | $1^5P_2$ | 2 | 1 | $2^{--}$ | 6642 | 6608 | 34 | $J/\psi(1S)\chi_{c1}(1P)$ |
| | | | | | | 6653 | −11 | $J/\psi(1S)\chi_{c2}(1P)$ |
| $T_{(4c)3^{--}}(6654)(1P)$ | $1^5P_3$ | 2 | 1 | $3^{--}$ | 6654 | 6653 | 1 | $J/\psi(1S)\chi_{c2}(1P)$ |
| $T_{(4c)0^{++}}(6711)(2S)$ | $2^1S_0$ | 0 | 0 | $0^{++}$ | 6711 | 5968 | 743 | $\eta_c(1S)\eta_c(1S)$ |
| | | | | | | 6194 | 517 | $J/\psi(1S)J/\psi(1S)$ |
| $T_{(4c)1^{+-}}(6734)(2S)$ | $2^3S_1$ | 1 | 0 | $1^{+-}$ | 6734 | 6081 | 653 | $\eta_c(1S)J/\psi(1S)$ |
| $T_{(4c)2^{++}}(6781)(2S)$ | $3^5S_2$ | 2 | 0 | $2^{++}$ | 6781 | 6194 | 540 | $J/\psi(1S)J/\psi(1S)$ |
| $T_{(4c)2^{++}}(6887)(1D)$ | $1^1D_2$ | 0 | 2 | $2^{++}$ | 6887 | 6194 | 693 | $J/\psi(1S)J/\psi(1S)$ |
| $T_{(4c)1^{--}}(6869)(1D)$ | $1^3D_1$ | 1 | 2 | $1^{+-}$ | 6869 | 6081 | 788 | $\eta_c(1S)J/\psi(1S)$ |
| $T_{(4c)2^{+-}}(6885)(1D)$ | $1^3D_2$ | 1 | 2 | $2^{+-}$ | 6885 | 6808 | 77 | $\eta_c(1S)\psi_2(3823)$ |
| $T_{(4c)3^{+-}}(6901)(1D)$ | $1^3D_3$ | 1 | 2 | $3^{+-}$ | 6901 | 6827 | 74 | $\eta_c(1S)\psi_3(3842)$ |
| $T_{(4c)0^{++}}(6816)(1D)$ | $1^5D_0$ | 2 | 2 | $0^{++}$ | 6816 | 5968 | 848 | $\eta_c(1S)\eta_c(1S)$ |
| | | | | | | 6194 | 622 | $J/\psi(1S)J/\psi(1S)$ |
| $T_{(4c)1^{++}}(6843)(1D)$ | $1^5D_1$ | 2 | 2 | $1^{++}$ | 6843 | 6194 | 649 | $J/\psi(1S)J/\psi(1S)$ |
| $T_{(4c)2^{++}}(6857)(1D)$ | $1^5D_2$ | 2 | 2 | $2^{++}$ | 6857 | 6194 | 663 | $J/\psi(1S)J/\psi(1S)$ |
| $T_{(4c)3^{++}}(6895)(1D)$ | $1^5D_3$ | 2 | 2 | $3^{++}$ | 6895 | 6921 | −26 | $J/\psi(1S)\psi_2(3823)$ |
| | | | | | | 6940 | −45 | $J/\psi(1S)\psi_3(3842)$ |
| $T_{(4c)4^{++}}(6918)(1D)$ | $1^5D_4$ | 1 | 2 | $3^{+-}$ | 6918 | 6940 | −22 | $J/\psi(1S)\psi_3(3842)$ |
| $T_{(4c)1^{--}}(7052)(2P)$ | $2^1P_1$ | 0 | 1 | $1^{--}$ | 7052 | 6509 | 543 | $\eta_c(1S)h_c(1P)$ |
| $T_{(4c)0^{-+}}(7067)(2P)$ | $2^3P_0$ | 1 | 1 | $0^{-+}$ | 7067 | 6399 | 668 | $\eta_c(1S)\chi_{c0}(1P)$ |
| $T_{(4c)1^{-+}}(7064)(2P)$ | $2^3P_1$ | 1 | 1 | $1^{-+}$ | 7064 | 6495 | 569 | $\eta_c(1S)\chi_{c1}(1P)$ |
| $T_{(4c)2^{-+}}(7062)(2P)$ | $2^3P_2$ | 1 | 1 | $2^{-+}$ | 7062 | 6540 | 522 | $\eta_c(1S)\chi_{c2}(1P)$ |
| $T_{(4c)1^{--}}(7098)(2P)$ | $2^5P_1$ | 2 | 1 | $1^{--}$ | 7098 | 6509 | 589 | $\eta_c(1S)h_c(1P)$ |

*(Table continued)*

TABLE V. *(Continued)*

| State | $n^{2S+1}L_J$ | $S$ | $L$ | $J^{PC}$ | $M_{cc\bar{c}\bar{c}}$ | $M_{\text{thr}}$ | $\Delta$ | Meson pair |
|---|---|---|---|---|---|---|---|---|
| $T_{(4c)2^{--}}(7097)(2P)$ | $2^5P_2$ | 2 | 1 | $2^{--}$ | 7097 | 6608 | | $J/\psi(1S)\chi_{c1}(1P)$ |
| $T_{(4c)3^{--}}(7094)(2P)$ | $2^5P_3$ | 2 | 1 | $3^{--}$ | 7094 | 6653 | | $J/\psi(1S)\chi_{c2}(1P)$ |
| $T_{(4c)0^{++}}(7189)(3S)$ | $3^1S_0$ | 0 | 0 | $0^{++}$ | 7189 | 5968 | 1221 | $\eta_c(1S)\eta_c(1S)$ |
| | | | | | | 6194 | 995 | $J/\psi(1S)J/\psi(1S)$ |
| $T_{(4c)1^{+-}}(7218)(3S)$ | $3^3S_1$ | 1 | 0 | $1^{+-}$ | 7218 | 6081 | 1137 | $\eta_c(1S)J/\psi(1S)$ |
| $T_{(4c)2^{++}}(7251)(3S)$ | $3^5S_2$ | 2 | 0 | $2^{++}$ | 7251 | 6194 | 1057 | $J/\psi(1S)J/\psi(1S)$ |

TABLE VI. Our predicted masses of ground and excited states of $bb\bar{b}\,\bar{b}$ tetraquark composed from axial-vector diquarks and the corresponding meson-meson thresholds. $\Delta = M_{bc\bar{b}\,\bar{c}} - M_{\text{thr}}$. Masses are given in unit of MeV.

| State | $n^{2S+1}L_J$ | $S$ | $L$ | $J^{PC}$ | $M_{bb\bar{b}\bar{b}}$ | $M_{\text{thr}}$ | $\Delta$ | Meson pair |
|---|---|---|---|---|---|---|---|---|
| $T_{(4b)0^{++}}(19286)(1S)$ | $1^1S_0$ | 0 | 0 | $0^{++}$ | 19286 | 18798 | 488 | $\eta_b(1S)\eta_b(1S)$ |
| | | | | | | 18921 | 365 | $\Upsilon(1S)\Upsilon(1S)$ |
| $T_{(4b)1^{+-}}(19292)(1S)$ | $1^3S_1$ | 1 | 0 | $1^{+-}$ | 19292 | 18859 | 433 | $\eta_b(1S)\Upsilon(1S)$ |
| $T_{(4b)2^{++}}(19302)(1S)$ | $1^5S_2$ | 2 | 0 | $2^{++}$ | 19302 | 18921 | 381 | $\Upsilon(1S)\Upsilon(1S)$ |
| | | | | | 19508 | 19298 | 210 | $\eta_b(1S)h_b(1P)$ |
| $T_{(4b)1^{--}}(19508)(1P)$ | $1^1P_1$ | 0 | 1 | $1^{--}$ | | 19320 | 188 | $\Upsilon(1S)\chi_{b0}(1P)$ |
| | | | | | | 19353 | 155 | $\Upsilon(1S)\chi_{b0}(1P)$ |
| | | | | | | 19373 | 135 | $\Upsilon(1S)\chi_{b0}(1P)$ |
| $T_{(4b)0^{-+}}(19505)(1P)$ | $1^3P_0$ | 1 | 1 | $0^{-+}$ | 19505 | 19258 | 247 | $\eta_b(1S)\chi_{b0}(1P)$ |
| | | | | | | 19360 | 145 | $\Upsilon(1S)h_b(1P)$ |
| $T_{(4b)1^{-+}}(19507)(1P)$ | $1^3P_1$ | 1 | 1 | $1^{-+}$ | 19507 | 19291 | 216 | $\eta_b(1S)\chi_{b1}(1P)$ |
| | | | | | | 19360 | 147 | $\Upsilon(1S)h_b(1P)$ |
| $T_{(4b)2^{-+}}(19511)(1P)$ | $1^3P_2$ | 1 | 1 | $2^{-+}$ | 19511 | 19311 | 200 | $\eta_b(1S)\chi_{b2}(1P)$ |
| | | | | | | 19360 | 151 | $\Upsilon(1S)h_b(1P)$ |
| | | | | | | 19298 | 208 | $\eta_b(1S)h_b(1P)$ |
| $T_{(4b)1^{--}}(19506)(1P)$ | $1^5P_1$ | 2 | 1 | $1^{--}$ | 19506 | 19320 | 186 | $\Upsilon(1S)\chi_{b0}(1P)$ |
| | | | | | | 19353 | 153 | $\Upsilon(1S)\chi_{b1}(1P)$ |
| | | | | | | 19373 | 133 | $\Upsilon(1S)\chi_{b2}(1P)$ |
| $T_{(4b)2^{--}}(19510)(1P)$ | $1^5P_2$ | 2 | 1 | $2^{--}$ | 19510 | 19353 | 157 | $\Upsilon(1S)\chi_{b1}(1P)$ |
| | | | | | | 19373 | 137 | $\Upsilon(1S)\chi_{b2}(1P)$ |
| $T_{(4b)3^{--}}(19517)(1P)$ | $1^5P_3$ | 2 | 1 | $3^{--}$ | 19517 | 19373 | 144 | $\Upsilon(1S)\chi_{b2}(1P)$ |
| $T_{(4b)0^{++}}(19571)(2S)$ | $2^1S_0$ | 0 | 0 | $0^{++}$ | 19571 | 18798 | 773 | $\eta_b(1S)\eta_b(1S)$ |
| | | | | | | 18921 | 650 | $\Upsilon(1S)\Upsilon(1S)$ |
| $T_{(4b)1^{+-}}(19576)(2S)$ | $2^3S_1$ | 1 | 0 | $1^{+-}$ | 19576 | 18859 | 717 | $\eta_b(1S)\Upsilon(1S)$ |
| $T_{(4b)2^{++}}(19581)(2S)$ | $2^5S_2$ | 2 | 0 | $2^{++}$ | 19581 | 18921 | 660 | $\Upsilon(1S)\Upsilon(1S)$ |
| $T_{(4b)2^{++}}(19582)(1D)$ | $1^1D_2$ | 0 | 2 | $2^{++}$ | 19685 | 18921 | 764 | $\Upsilon(1S)\Upsilon(1S)$ |
| $T_{(4b)1^{+-}}(19680)(1D)$ | $1^3D_1$ | 1 | 2 | $1^{+-}$ | 19680 | 18859 | 821 | $\eta_b(1S)\Upsilon(1S)$ |
| $T_{(4b)2^{+-}}(19684)(1D)$ | $1^3D_2$ | 1 | 2 | $2^{+-}$ | 19684 | 19562 | 122 | $\eta_b(1S)\Upsilon_2(1D)$ |
| $T_{(4b)3^{+-}}(19693)(1D)$ | $1^3D_3$ | 1 | 2 | $3^{+-}$ | 19693 | 19812 | −119 | $h_b(1P)\chi_{b2}(1P)$ |

*(Table continued)*

TABLE VI. *(Continued)*

| State | $n^{2S+1}L_J$ | $S$ | $L$ | $J^{PC}$ | $M_{bb\bar{b}\bar{b}}$ | $M_{\rm thr}$ | $\Delta$ | Meson pair |
|---|---|---|---|---|---|---|---|---|
| $T_{(4b)0^{++}}(19675)(1D)$ | $1^5D_0$ | 2 | 2 | $0^{++}$ | 19675 | 18798<br>18921 | 877<br>754 | $\eta_b(1S)\eta_b(1S)$<br>$\Upsilon(1S)\Upsilon(1S)$ |
| $T_{(4b)1^{++}}(19677)(1D)$ | $1^5D_1$ | 2 | 2 | $1^{++}$ | 19677 | 18921 | 756 | $\Upsilon(1S)\Upsilon(1S)$ |
| $T_{(4b)2^{++}}(19681)(1D)$ | $1^5D_2$ | 2 | 2 | $2^{++}$ | 19681 | 18921 | 760 | $\Upsilon(1S)\Upsilon(1S)$ |
| $T_{(4b)3^{++}}(19687)(1D)$ | $1^5D_3$ | 2 | 2 | $3^{++}$ | 19687 | 19624 | 63 | $\Upsilon(1S)\Upsilon_2(1D)$ |
| $T_{(4b)4^{++}}(19694)(1D)$ | $1^5D_4$ | 2 | 2 | $4^{++}$ | 19694 | 19824 | −130 | $\chi_{b2}(1P)\chi_{b2}(1P)$ |
| $T_{(4b)1^{--}}(19790)(2P)$ | $2^1P_1$ | 0 | 1 | $1^{--}$ | 19790 | 19298 | 492 | $\eta_b(1S)h_b(1P)$ |
| $T_{(4b)0^{-+}}(19791)(2P)$ | $2^3P_0$ | 1 | 1 | $0^{-+}$ | 19791 | 19258 | 533 | $\eta_b(1S)\chi_{b0}(1P)$ |
| $T_{(4b)1^{-+}}(19791)(2P)$ | $2^3P_1$ | 1 | 1 | $1^{-+}$ | 19791 | 19291 | 500 | $\eta_b(1S)\chi_{b1}(1P)$ |
| $T_{(4b)2^{-+}}(19792)(2P)$ | $2^3P_2$ | 1 | 1 | $2^{-+}$ | 19792 | 19311 | 481 | $\eta_b(1S)\chi_{b2}(1P)$ |
| $T_{(4b)1^{--}}(19793)(2P)$ | $2^5P_1$ | 2 | 1 | $1^{--}$ | 19793 | 19298 | 495 | $\eta_b(1S)h_b(1P)$ |
| $T_{(4b)2^{--}}(19793)(2P)$ | $2^5P_2$ | 2 | 1 | $2^{--}$ | 19793 | 19353 | 440 | $\Upsilon(1S)\chi_{b1}(1P)$ |
| $T_{(4b)3^{--}}(19794)(2P)$ | $2^5P_3$ | 2 | 1 | $3^{--}$ | 19794 | 19373 | 421 | $\Upsilon(1S)\chi_{b2}(1P)$ |
| $T_{(4b)0^{++}}(19911)(3S)$ | $3^1S_0$ | 0 | 0 | $0^{++}$ | 19911 | 18798<br>18921 | 1113<br>990 | $\eta_b(1S)\eta_b(1S)$<br>$\Upsilon(1S)\Upsilon(1S)$ |
| $T_{(4b)1^{+-}}(19913)(3S)$ | $3^3S_1$ | 1 | 0 | $1^{+-}$ | 19913 | 18859 | 1054 | $\eta_b(1S)\Upsilon(1S)$ |
| $T_{(4b)2^{++}}(19917)(3S)$ | $3^5S_2$ | 2 | 0 | $2^{++}$ | 19917 | 18921 | 996 | $\Upsilon(1S)\Upsilon(1S)$ |

close to their respective $\Upsilon(1S)\Upsilon_2(1D)$ and $\chi_{b2}(1P)\chi_{b2}(1P)$ thresholds, with $\Delta \approx 63$ and $-130$ MeV. Following the arguments of Faustov *et al.* [65], such near-threshold or subthreshold states could be narrow due to suppressed decays, offering a more promising, though experimentally challenging, avenue for discovery in channels other than $\Upsilon(1S)\Upsilon(1S)$.

The mixed-flavor $bc\bar{b}\bar{c}$ system presents an intermediate case with distinct challenges. All states in Tables VII and VIII are predicted to be above the lowest thresholds [e.g., $\eta_b(1S)\eta_c(1S)$], with $\Delta$ values of 300–400 MeV for the ground states, escalating beyond 1 GeV for radial excitations. Consequently, these tetraquarks are also expected to be broad resonances decaying into combinations of $B_c$, charmonium, and bottomonium mesons. The absence of any observed structure in this mass region is consistent with this prediction. Moreover, the experimental signature is more complex than for fully charm or fully bottom systems, as the final states lack the distinctive back-to-back dilepton pairs, further complicating detection.

Our comprehensive analysis allows us to propose specific candidate assignments for the observed exotic $X$ states, as summarized in Table IX. The assignments are based on the agreement between our calculated masses and the experimental measurements, combined with the decay dynamics inferred from the $\Delta$ values. The $X(6200)$, reported by ATLAS with a mass of $6220 \pm 50^{+40}_{-50}$ MeV, is identified as the narrow, subthreshold $T_{(4c)0^{++}}(1S)$ state at 6162 MeV. Its negative $\Delta$ relative to the di-$J/\psi$ channel suppresses the fall-apart decay, consistent with it being a potentially narrow resonance, an interpretation also favored in other works [65,71]. The broad enhancement denoted as $X(6400)$ by LHCb is a natural candidate for our first radial excitations of the $2^{++}$ configuration, specifically the $T_{(4c)2^{++}}(1S)$ states at 6343 and 6339 MeV. These states, lying significantly above the di-$J/\psi$ threshold ($\Delta \sim 150$ MeV), are expected to be broad, aligning with the experimental observation. For the prominent $X(6600)$ structure (CMS: $6552 \pm 10 \pm 12$ MeV; ATLAS: $6620 \pm 30$ MeV), we propose a multiplet of $1P$ states—including the $T_{(4c)0^{-+}}(6623)(1P)$, $T_{(4c)1^{-+}}(6629)(1P)$, and $T_{(4c)2^{-+}}(6638)(1P)$ whose small positive $\Delta$ values relative to channels like $J/\psi(1S)h_c(1P)$ support a broad width. This assignment of $X(6600)$ to $P$-wave states is a common feature in several past theoretical interpretations [37,65]. The well-established $X(6900)$ resonance is interpreted as a dense overlapping band of $1D$-wave states. Our candidates span masses from 6869 to 6918 MeV, covering the $J^{PC}$ quantum numbers $1^{--}$, $2^{++}$, $3^{+-}$, $1^{++}$, $3^{++}$, and $4^{++}$, which collectively match the measured mass of ∼6905 MeV. The $X(7200)/X(7300)$ structures are associated with the radially excited $3S$ states: the $T_{(4c)0^{++}}(7189)(3S)$ and the $T_{(4c)2^{++}}(7251)(3S)$. These assignments are consistent with the pattern identified in Ref. [37], where the higher-mass structures are linked to radial excitations. Furthermore, a recent angular analysis by the CMS Collaboration [73] suggests that the most probable quantum numbers for the $X(6900)$ state are $J^{PC} = 2^{++}$, which aligns with several of our predicted $1D$-wave states, such as the $T_{(4c)2^{++}}(1D)$ at 6887 MeV. This collective interpretation, which maps the observed resonances to specific radial, orbital, and spin excitations within the diquark-antidiquark model employing a color Coulomb plus nonrelativistic screened potential, successfully confronts the current LHC data and provides a coherent framework for understanding the spectroscopy of fully heavy tetraquarks.

TABLE VII. Calculated masses for $1S$, $1P$, and $2S$ states of $bc\bar{b}\,\bar{c}$ tetraquark composed from axial-vector diquarks and the corresponding meson-meson thresholds. $\Delta = M_{bc\bar{b}\,\bar{c}} - M_{\mathrm{thr}}$. Masses are given in the unit of MeV.

| State | $n^{2S+1}L_J$ | $S$ | $L$ | $J^{PC}$ | $M_{bc\bar{b}\,\bar{c}}$ | $M_{\mathrm{thr}}$ | $\Delta$ | Meson pair |
|---|---|---|---|---|---|---|---|---|
| $T_{(2bc)0^{++}}(12810)(1S)$ | $1^1S_0$ | 0 | 0 | $0^{++}$ | 12810 | 12383 | 427 | $\eta_c(1S)\eta_b(1S)$ |
| $T_{(2bc)1^{+-}}(12827)(1S)$ | $1^3S_1$ | 1 | 0 | $1^{+-}$ | 12827 | 12444 | 383 | $\eta_c(1S)\Upsilon(1S)$ |
| $T_{(2bc)2^{++}}(12855)(1S)$ | $1^5S_2$ | 2 | 0 | $2^{++}$ | 12855 | 12557 | 298 | $J/\psi(1S)\Upsilon(1S)$ |
| $T_{(2bc)1^{--}}(13075)(1P)$ | $1^1P_1$ | 0 | 1 | $1^{--}$ | 13075 | 12875 | 200 | $\chi_{c0}(1P)\Upsilon(1S)$ |
| | | | | | | 12883 | 192 | $\eta_c(1S)h_b(1P)$ |
| | | | | | | 12924 | 151 | $h_c(1P)\eta_b(1S)$ |
| | | | | | | 12956 | 119 | $J/\psi(1S)\chi_{b0}(1P)$ |
| | | | | | | 12971 | 104 | $\chi_{c1}(1P)\Upsilon(1S)$ |
| | | | | | | 12990 | 85 | $J/\psi(1S)\chi_{b1}(1P)$ |
| | | | | | | 13009 | 66 | $J/\psi(1S)\chi_{b2}(1P)$ |
| | | | | | | 13016 | 59 | $\chi_{c2}(1P)\Upsilon(1S)$ |
| $T_{(2bc)0^{-+}}(13072)(1P)$ | $1^3P_0$ | 1 | 1 | $0^{-+}$ | 13072 | 12813 | 259 | $\chi_{c0}(1P)\eta_b(1S)$ |
| | | | | | | 12843 | 229 | $\eta_c(1S)\chi_{b0}(1P)$ |
| | | | | | | 12986 | 86 | $h_c(1P)\Upsilon(1S)$ |
| | | | | | | 12996 | 76 | $J/\psi(1S)h_b(1P)$ |
| $T_{(2bc)1^{-+}}(13075)(1P)$ | $1^3P_1$ | 1 | 1 | $1^{-+}$ | 13075 | 12877 | 198 | $\eta_c(1P)\chi_{b1}(1P)$ |
| | | | | | | 12909 | 166 | $\chi_{c1}(1P)\eta_b(1S)$ |
| | | | | | | 12986 | 89 | $h_c(1P)\Upsilon(1S)$ |
| | | | | | | 12996 | 79 | $J/\psi(1S)h_b(1P)$ |
| $T_{(2bc)2^{-+}}(13080)(1P)$ | $1^3P_2$ | 1 2 | 1 1 | $2^{-+}$ | 13080 | 12896 | 184 | $\eta_c(1S)\chi_{b2}(1P)$ |
| | | | | | | 12955 | 125 | $\chi_{c2}(1P)\eta_b(1S)$ |
| | | | | | | 12986 | 94 | $h_c(1P)\Upsilon(1S)$ |
| | | | | | | 12996 | 84 | $J/\psi(1S)h_b(1P)$ |
| $T_{(2bc)1^{--}}(13075)(1P)$ | $1^5P_1$ | | | $1^{--}$ | 13075 | 12875 | 200 | $\chi_{c0}(1P)\Upsilon(1S)$ |
| | | | | | | 12883 | 192 | $\eta_c(1S)h_b(1P)$ |
| | | | | | | 12924 | 151 | $h_c(1P)\eta_b(1S)$ |
| | | | | | | 12954 | 121 | $J/\psi(1S)\chi_{b0}(1P)$ |
| | | | | | | 12971 | 104 | $\chi_{c1}(1P)\Upsilon(1S)$ |
| | | | | | | 12990 | 85 | $J/\psi(1S)\chi_{b1}(1P)$ |
| | | | | | | 13009 | 66 | $J/\psi(1S)\chi_{b2}(1P)$ |
| | | | | | | 13016 | 59 | $\chi_{c1}(1P)\Upsilon(1S)$ |
| $T_{(2bc)2^{--}}(13082)(1P)$ | $1^5P_2$ | 2 | 1 | $2^{--}$ | 13082 | 12971 | 111 | $\chi_{c1}(1P)\Upsilon(1S)$ |
| | | | | | | 12990 | 99 | $J/\psi(1S)\chi_{b1}(1P)$ |
| | | | | | | 13009 | 73 | $J/\psi(1S)\chi_{b2}(1P)$ |
| | | | | | | 13016 | 66 | $\chi_{c2}(1P)\Upsilon(1S)$ |
| $T_{(2bc)3^{--}}(13088)(1P)$ | $1^5P_3$ | 2 | 1 | $3^{--}$ | 13088 | 13009 | 79 | $J/\psi(1S)\chi_{b2}(1P)$ |
| | | | | | | 13016 | 72 | $\chi_{c2}(1P)\Upsilon(1S)$ |
| $T_{(2bc)0^{++}}(13236)(2S)$ | $2^1S_0$ | 0 | 0 | $0^{++}$ | 13236 | 12383 | 853 | $\eta_c(1S)\eta_b(1S)$ |
| $T_{(2bc)1^{+-}}(13244)(2S)$ | $2^3S_1$ | 1 | 0 | $1^{+-}$ | 13244 | 12444 | 800 | $\eta_c(1S)\Upsilon(1S)$ |
| $T_{(2bc)2^{++}}(13261)(2S)$ | $2^5S_2$ | 2 | 0 | $2^{++}$ | 13261 | 12557 | 704 | $J/\psi(1S)\Upsilon(1S)$ |

TABLE VIII. Calculated masses of $1D$, $2P$, and $3S$ states of $bc\bar{b}\,\bar{c}$ tetraquark composed from axial-vector diquarks and the corresponding meson-meson thresholds. $\Delta = M_{bc\bar{b}\,\bar{c}} - M_{\rm thr}$ Masses are given in unit of MeV.

| State | $n^{2S+1}L_J$ | $S$ | $L$ | $J^{PC}$ | $M_{bc\bar{b}\,\bar{c}}$ | $M_{\rm thr}$ | $\Delta$ | Meson pair |
|---|---|---|---|---|---|---|---|---|
| $T_{(2bc)2^{++}}(13278)(1D)$ | $1^1D_2$ | 0 | 2 | $2^{++}$ | 13278 | 12557 | 721 | $J/\psi(1S)\Upsilon(1S)$ |
| $T_{(2bc)1^{+-}}(13272)(1D)$ | $1^3D_1$ | 1 | 2 | $1^{+-}$ | 13272 | 12444 | 828 | $\eta_c(1S)\Upsilon(1S)$ |
| $T_{(2bc)2^{+-}}(13276)(1D)$ | $1^3D_2$ | 1 | 2 | $2^{+-}$ | 13276 | 13148 | 128 | $\eta_c(1S)\Upsilon_2(1D)$ |
| | | | | | | 13222 | 54 | $\psi_2(3823)\eta_b(1S)$ |
| $T_{(2bc)3^{+-}}(13301)(1D)$ | $1^3D_3$ | 1 | 2 | $3^{+-}$ | 13301 | 13241 | 60 | $\psi_3(3842)\eta_b(1S)$ |
| $T_{(2bc)0^{++}}(13287)(1D)$ | $1^5D_0$ | 2 | 2 | $0^{++}$ | 13287 | 12383 | 904 | $\eta_c(1S)\eta_b(1S)$ |
| $T_{(2bc)1^{++}}(13284)(1D)$ | $2^5D_1$ | 2 | 2 | $1^{++}$ | 13284 | 12557 | 727 | $J/\psi(1S)\Upsilon(1S)$ |
| $T_{(2bc)2^{++}}(13297)(1D)$ | $2^5D_2$ | 2 | 2 | $2^{++}$ | 13297 | 12557 | 740 | $J/\psi(1S)\Upsilon(1S)$ |
| $T_{(2bc)3^{++}}(13298)(1D)$ | $1^5D_3$ | 2 | 2 | $3^{++}$ | 13298 | 13261 | 37 | $J/\psi(1S)\Upsilon_2(1D)$ |
| | | | | | | 13284 | 14 | $\psi_2(3823)\Upsilon(1S)$ |
| | | | | | | 13303 | −5 | $\psi_3(3842)\Upsilon(1S)$ |
| $T_{(2bc)1^{--}}(13398)(2P)$ | $2^1P_1$ | 0 | 1 | $1^{--}$ | 13398 | 12875 | 523 | $\chi_{c0}(1P)\Upsilon(1S)$ |
| $T_{(2bc)0^{-+}}(13404)(2P)$ | $2^3P_0$ | 1 | 1 | $0^{-+}$ | 13404 | 12813 | 591 | $\chi_{c0}(1P)\eta_b(1S)$ |
| $T_{(2bc)1^{-+}}(13404)(2P)$ | $2^3P_1$ | 1 | 1 | $1^{-+}$ | 13404 | 12877 | 527 | $\eta_c(1S)\chi_{b1}(1P)$ |
| $T_{(2bc)2^{-+}}(13404)(2P)$ | $2^3P_2$ | 1 | 1 | $2^{-+}$ | 13404 | 12896 | 508 | $\eta_c(1S)\chi_{b2}(1S)$ |
| $T_{(2bc)1^{--}}(13406)(2P)$ | $2^5P_1$ | 2 | 1 | $1^{--}$ | 13406 | 12875 | 531 | $\chi_{c0}(1P)\Upsilon(1S)$ |
| $T_{(2bc)2^{--}}(13407)(2P)$ | $2^5P_2$ | 2 | 1 | $2^{--}$ | 13407 | 12971 | 436 | $\chi_{c1}(1P)\Upsilon(1S)$ |
| $T_{(2bc)3^{--}}(13408)(2P)$ | $2^5P_3$ | 2 | 1 | $3^{--}$ | 13408 | 13009 | 399 | $J/\psi(1S)\chi_{b2}(1P)$ |
| $T_{(2bc)0^{++}}(13518)(3S)$ | $2^1S_0$ | 0 | 0 | $0^{++}$ | 13518 | 12383 | 1135 | $\eta_c(1S)\eta_b(1S)$ |
| $T_{(2bc)1^{+-}}(13515)(3S)$ | $2^3S_1$ | 1 | 0 | $1^{+-}$ | 13515 | 12444 | 1071 | $\eta_c(1S)\Upsilon(1S)$ |
| $T_{(2bc)2^{++}}(13533)(3S)$ | $2^5S_2$ | 2 | 0 | $2^{++}$ | 13533 | 12557 | 976 | $J/\psi(1S)\Upsilon(1S)$ |

Figures 4–12 present a comprehensive theoretical spectrum of $S$-, $P$-, and $D$-wave tetraquark states for fully charmed, mixed heavy, and fully bottom systems, with masses categorized by principal quantum number $n = \{1, 2, 3, 4\}$ and overlaid with experimentally observed resonances ($X$ states) in the charmoniumlike region. For the fully charmed system, several experimental states notably the $X(6900)$ and $X(7200)$ families clearly show proximity to predicted theoretical masses, while the mixed and bottomonium systems remain purely theoretical predictions at higher mass scales. The visual organization clearly distinguishes between different angular momentum states, excitation levels, and experimental collaborations, facilitating a direct comparison between theoretical predictions and recent collider data in the search for exotic multiquark hadrons.

### D. Comparison with previous theoretical results

In Tables X and XI, we compare our predicted mass spectra for fully heavy $cc\bar{c}\bar{c}$ and $bb\bar{b}\bar{b}$ tetraquark states with few representative results obtained using a variety of theoretical frameworks, including nonrelativistic and relativized quark models, potential models, QCD-inspired approaches, and diquark-antidiquark constructions. For the ground-state $S$-wave multiplets, our results are generally found to lie within the broad range of previous theoretical estimates. In particular, for the $1S$ states, our masses are systematically higher than those reported by Mistry and Majethiya [37], who employ a nonrelativistic potential model with different parameter choices, but are in very good agreement with the relativistic quark-diquark model of Faustov *et al.* [65]. Earlier calculations based on color-singlet clustering or effective diquark assumptions, such as those by Berezhnoy *et al.* [64], Lloyd and Vary [74], and Iwasaki [1], tend to span a wider mass range, reflecting the strong model dependence inherent in early treatments of multiquark dynamics.

For the radially excited $2S$ and $3S$ states, all approaches predict a relatively uniform upward mass shift of several hundred MeV relative to the ground state, with our results closely tracking the patterns observed in Refs. [75–77,81]. While some deviations at the level of 50–150 MeV are present, these differences can be attributed to variations in the assumed heavy-quark masses, confinement potentials, and spin-dependent interactions. Importantly, the internal splittings within each $nS$ multiplet ($0^{++}$, $1^{+-}$, $2^{++}$) are consistently reproduced across models, indicating a robust hierarchy driven by spin-spin interactions.

TABLE IX. The experimentally observed exotic $X$ states in the di-$J/\psi$ invariant mass spectrum, as reported by the LHCb, CMS, and ATLAS collaborations, alongside our corresponding theoretical tetraquark candidates. All masses are reported in the unit of MeV.

| | | | | | Possible candidates | | | | |
|---|---|---|---|---|---|---|---|---|---|
| Collaboration | State | Mass | Width | Refs. | State | $n^{2S+1}L_J$ | Spin | $J^{PC}$ | Mass |
| ATLAS | $X(6200)$ | $6220 \pm 50^{+40}_{-50}$ | $310 \pm 120^{+70}_{-80}$ | [15] | $T_{(4c)0^{++}}(1S)$ | $1^1S_0$ | 0 | $0^{++}$ | 6162 |
| LHCb | $X(6400)$ | $\approx 6400$ | broad | [13] | $T_{(4c)2^{++}}(1S)$ | $1^5S_2$ | 2 | $2^{++}$ | 6343 |
| CMS | $X(6600)$ | $6552 \pm 10 \pm 12$ | $124 \pm 29 \pm 34$ | [14] | $T_{(4c)0^{-+}}(1P)$ | $1^3P_0$ | 1 | $0^{-+}$ | 6623 |
| | | | | | $T_{(4c)1^{-+}}(1P)$ | $1^3P_1$ | 1 | $1^{-+}$ | 6629 |
| | | | | | $T_{(4c)2^{-+}}(1P)$ | $1^3P_2$ | 1 | $2^{-+}$ | 6638 |
| | | | | | $T_{(4c)1^{--}}(1P)$ | $1^5P_1$ | 2 | $1^{--}$ | 6630 |
| ATLAS | $X(6620)$ | $6620 \pm 30^{+20}_{-10}$ | $310 \pm 90^{+60}_{-110}$ | [15] | $T_{(4c)2^{--}}(1P)$ | $1^5P_2$ | 2 | $2^{--}$ | 6642 |
| LHCb | $X(6900)$ | $6905 \pm 11 \pm 7$ | $80 \pm 19 \pm 33$ | [13] | $T_{(4c)2^{++}}(1D)$ | $1^1D_2$ | 0 | $2^{++}$ | 6887 |
| | | | | | $T_{(4c)1^{--}}(1D)$ | $1^3D_1$ | 1 | $1^{--}$ | 6869 |
| | | | | | $T_{(4c)3^{+-}}(1D)$ | $1^3D_3$ | 1 | $3^{+-}$ | 6901 |
| | | | | | $T_{(4c)1^{++}}(1D)$ | $1^5D_1$ | 2 | $1^{++}$ | 6843 |
| | | | | | $T_{(4c)2^{++}}(1D)$ | $1^5D_2$ | 2 | $2^{++}$ | 6857 |
| | | | | | $T_{(4c)3^{++}}(1D)$ | $1^5D_3$ | 2 | $3^{++}$ | 6895 |
| CMS | $X(6886)$ | $6886 \pm 11 \pm 11$ | $168 \pm 33 \pm 69$ | [14] | $T_{(4c)4^{++}}(1D)$ | $1^5D_4$ | 2 | $4^{++}$ | 6918 |
| LHCb | $X(7200)$ | $\approx 7200$ | broad | [13] | $T_{(4c)0^{++}}(3S)$ | $3^1S_0$ | 0 | $0^{++}$ | 7189 |
| ATLAS | $X(7220)$ | $7220 \pm 30^{+20}_{-30}$ | $100^{+130+60}_{-70-50}$ | [15] | $T_{(4c)1^{+-}}(3S)$ | $3^3S_1$ | 1 | $1^{+-}$ | 7218 |
| CMS | $X(7300)$ | $7287 \pm 19 \pm 5$ | $95 \pm 46 \pm 20$ | [14] | $T_{(4c)2^{++}}(3S)$ | $3^5S_2$ | 2 | $2^{++}$ | 7251 |

A similar level of agreement is observed for the $P$- and $D$-wave sectors. For the $1P$ states, our predictions cluster near those of Karliner and Rosner [78], Tiwari *et al.* [76], and Debastiani and Navarra [75], with small systematic shifts that remain well within the theoretical uncertainties typical of orbital excitations. The near-degeneracy of several $P$-wave states with different total spin $S_T$ but the same orbital angular momentum $L_T$ is a common feature shared by all models, reflecting the relatively weaker role of spin-dependent forces in higher partial waves. For the $2P$ sector, a larger spread is visible, particularly when compared with Ref. [79], where two distinct parameter sets (Id and IId) lead to substantially different mass scales; nevertheless, our results align closely with the lower-mass (Id) solutions and with other quark-model predictions, suggesting that extreme parameter choices may overestimate the excitation energies. For the $D$-wave states, fewer results are available in the literature, but our predictions show excellent consistency with those of Karliner and Rosner [78] and Liu *et al.* [80], both in terms of absolute mass values and the ordering of the $J^{PC}$ multiplets. The same qualitative conclusions apply to the fully bottom $bb\bar{b}\bar{b}$ system, where all models predict a more compressed spectrum with smaller relative splittings, as expected from the heavier quark mass. Our results again lie between the lower predictions of Ref. [37] and the slightly higher relativistic estimates of Ref. [65], while remaining compatible with other potential-model calculations [64,78,80,82].

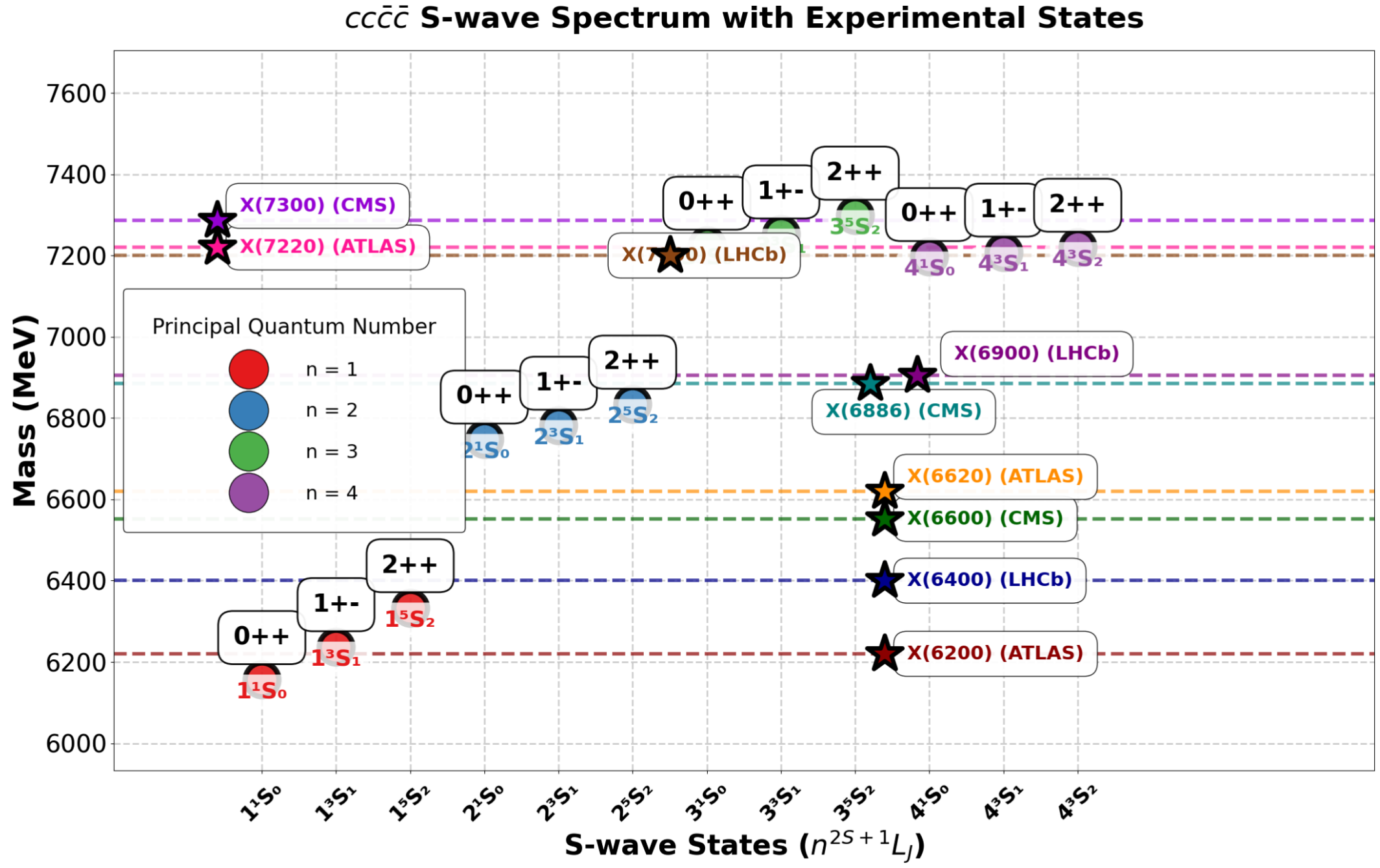

FIG. 4. $cc\bar{c}\bar{c}$ $S$-wave tetraquark spectrum with experimental states.

FIG. 5. $cc\bar{c}\,\bar{c}$ $P$-wave tetraquark spectrum with experimental states.

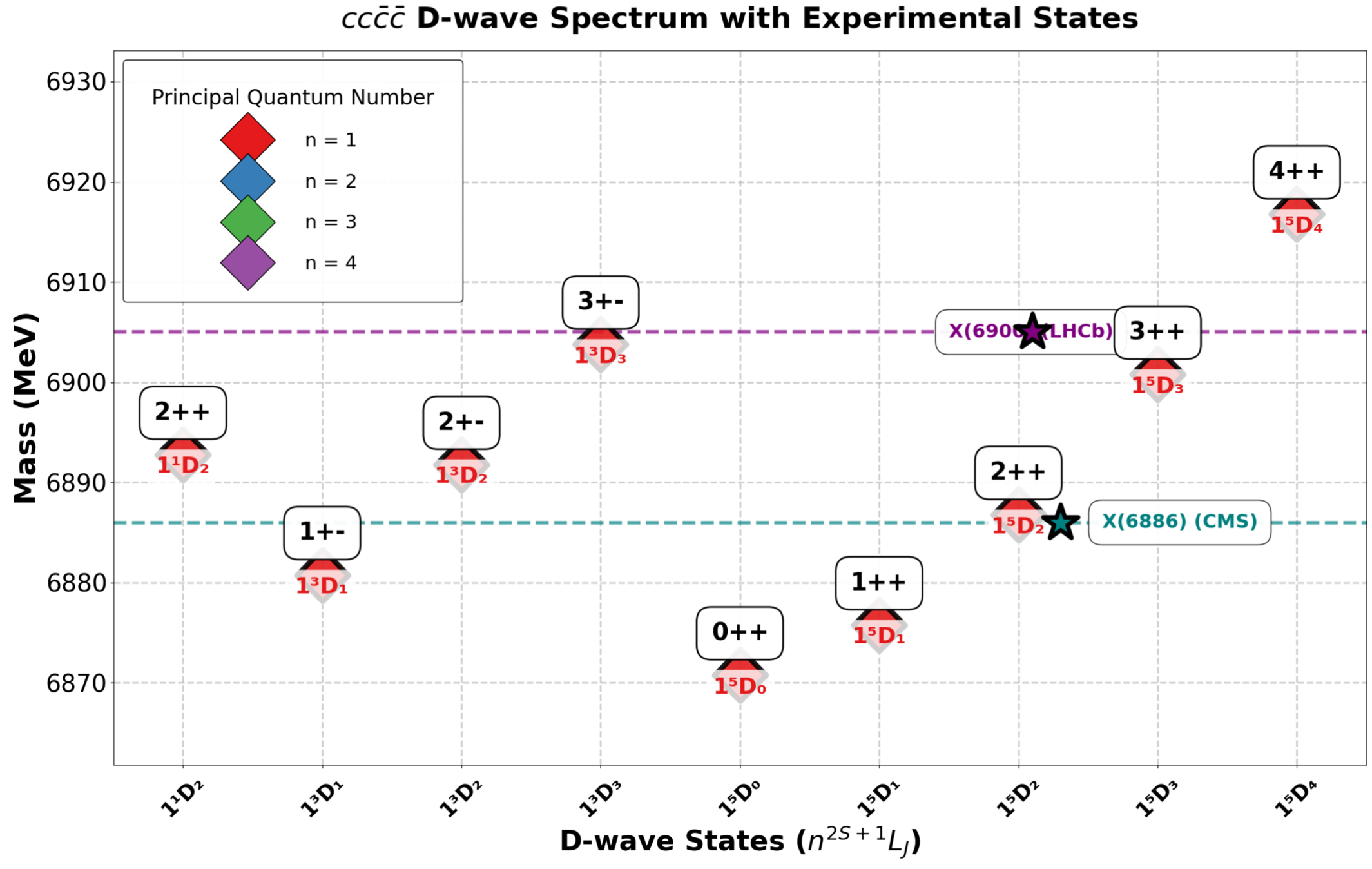

FIG. 6. $cc\bar{c}\,\bar{c}$ $D$-wave tetraquark spectrum with experimental states.

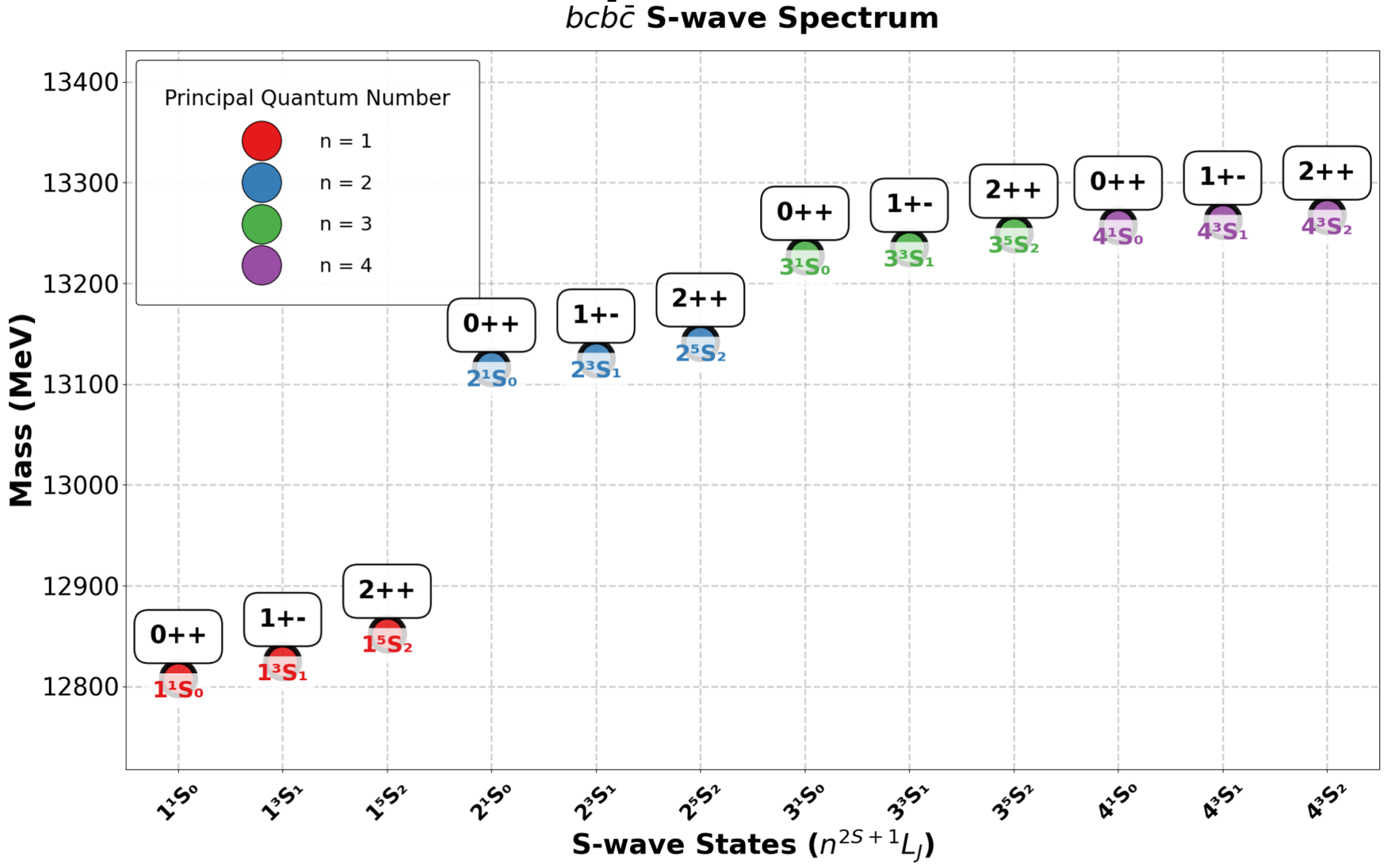

FIG. 7. $bc\bar{b}\,\bar{c}$ $S$-wave tetraquark spectrum.

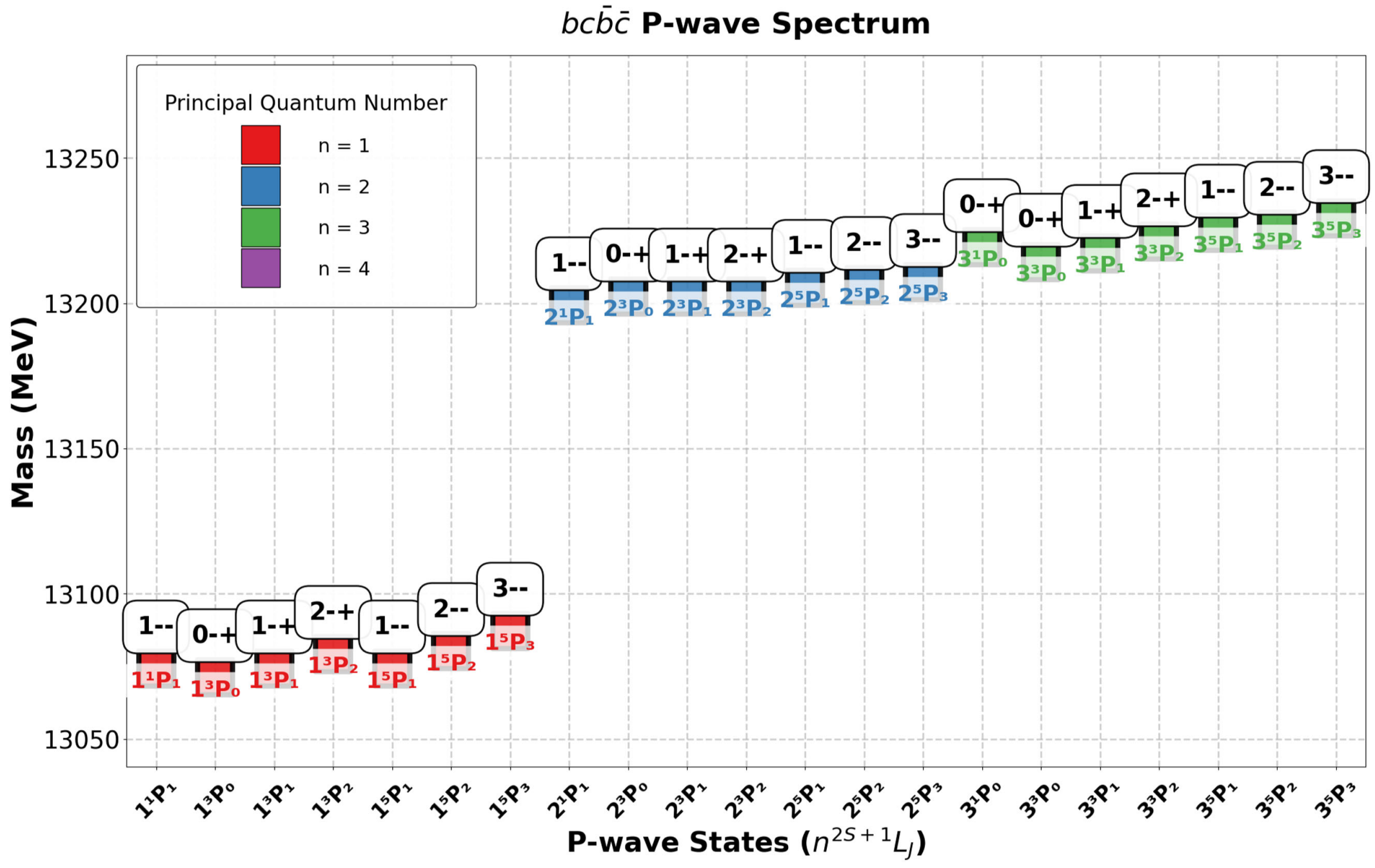

FIG. 8. $bc\bar{b}\,\bar{c}$ $P$-wave tetraquark spectrum.

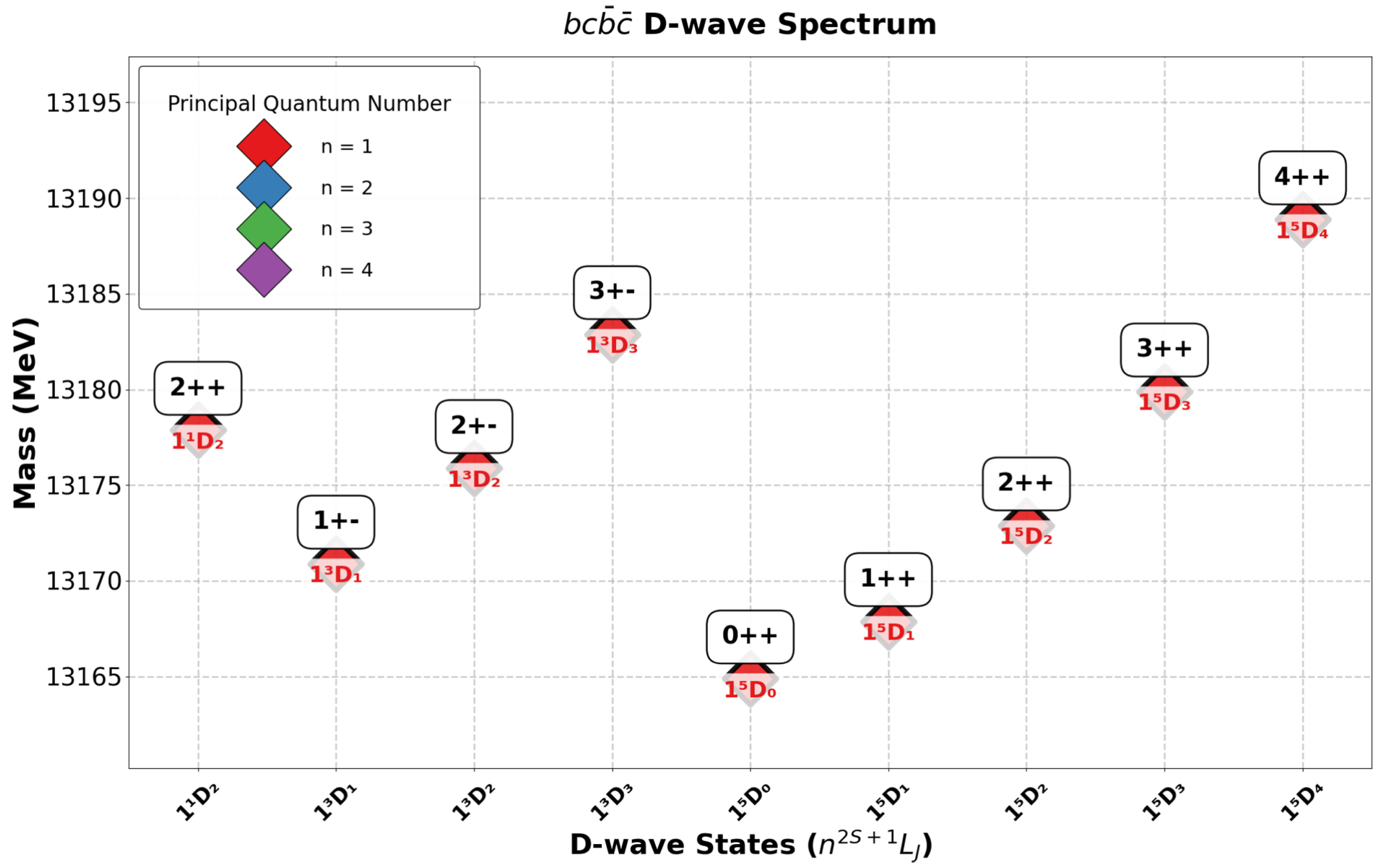

FIG. 9. $bc\bar{b}\bar{c}$ $D$-wave tetraquark spectrum.

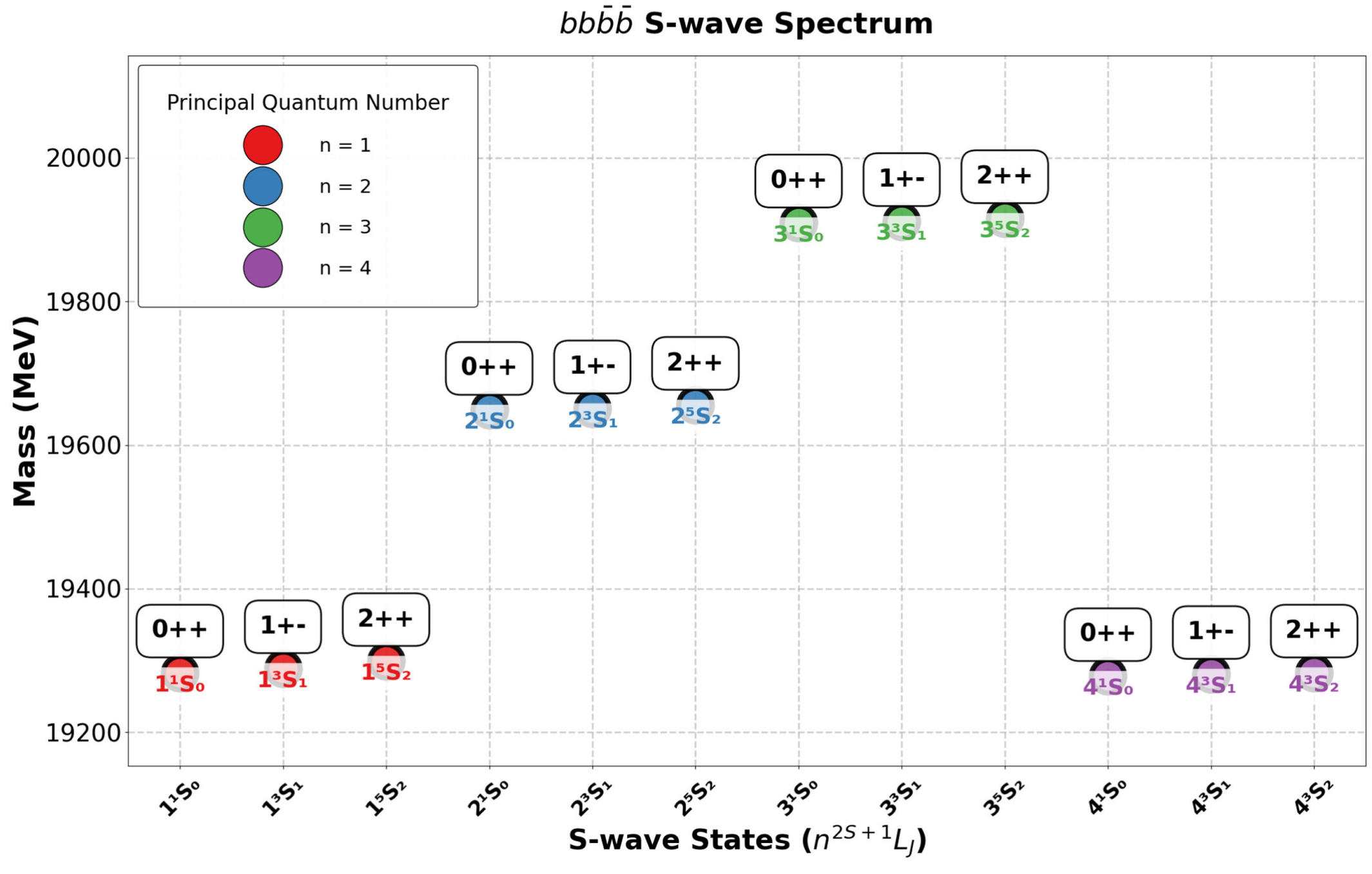

FIG. 10. $bb\bar{b}\bar{b}$ $S$-wave tetraquark spectrum.

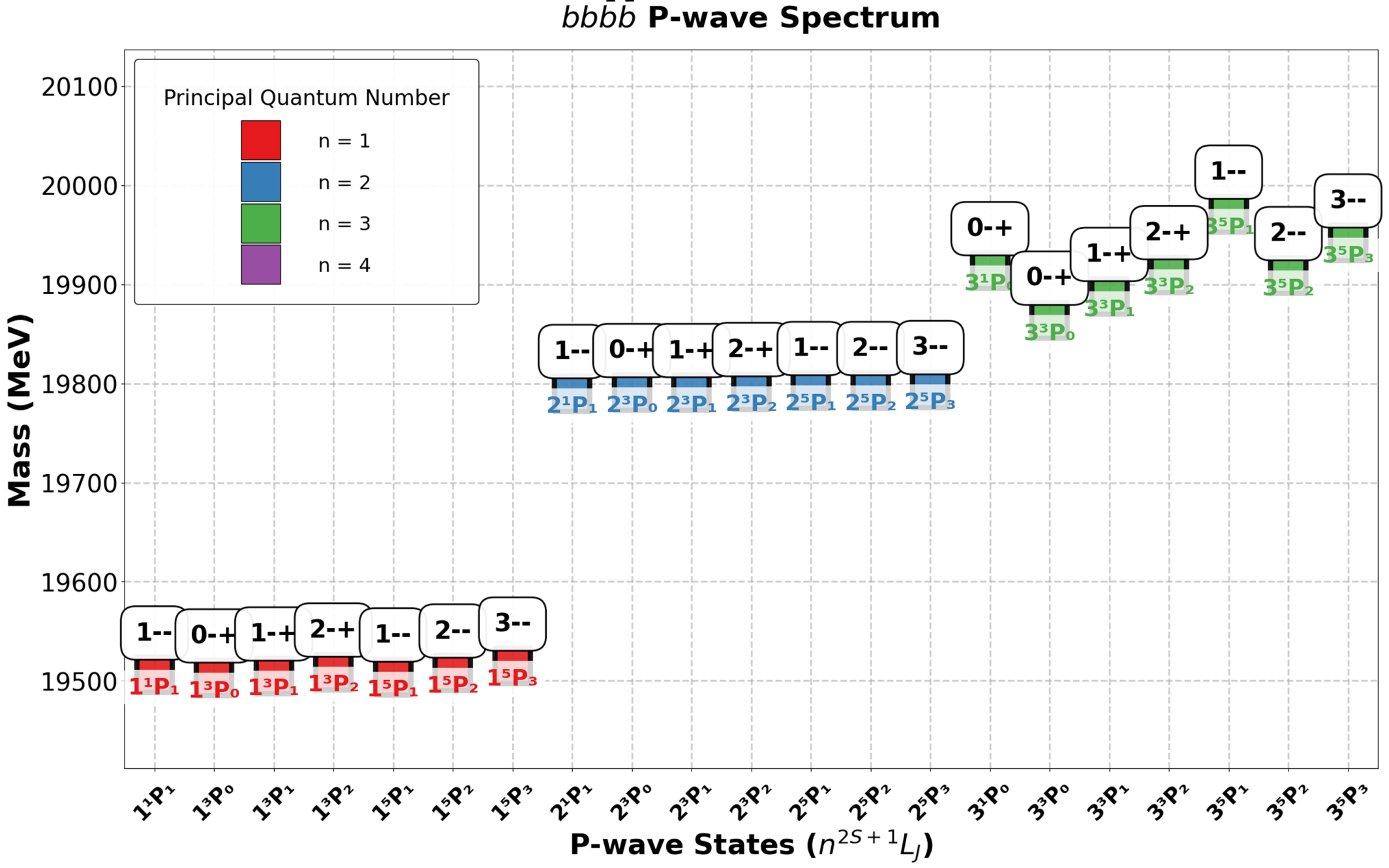

FIG. 11. $bb\bar{b}\bar{b}$ $P$-wave tetraquark spectrum.

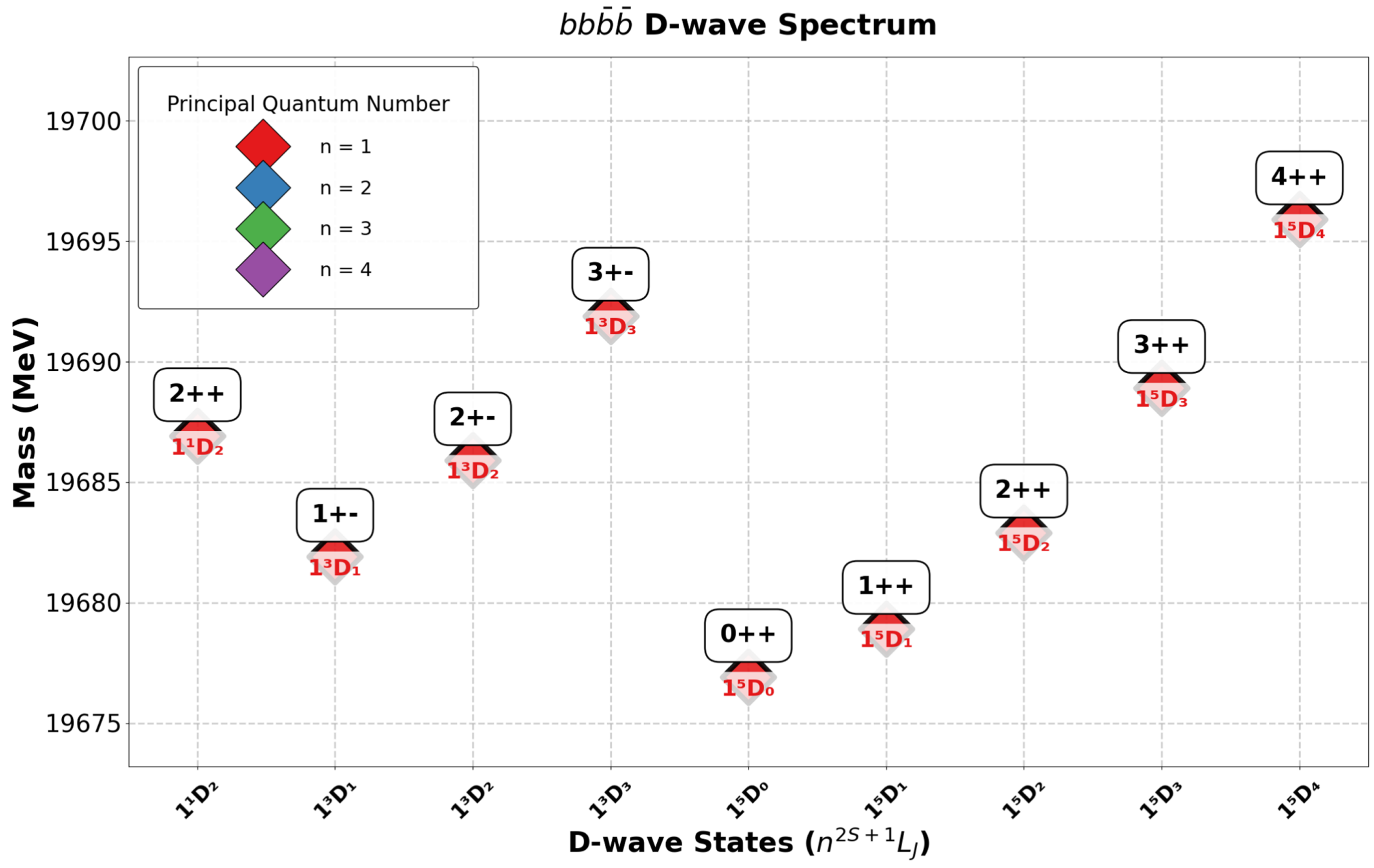

FIG. 12. $bb\bar{b}\bar{b}$ $D$-wave tetraquark spectrum.

Overall, despite the diversity of theoretical approaches and model assumptions, a clear global consistency emerges across the literature. Our predicted spectra follow the same systematic trends observed in previous studies, namely, regular radial and orbital excitation patterns, stable spin-multiplet splittings, and a reduced sensitivity to spin effects for higher $L$. This agreement strongly supports the reliability of our framework and confirms that our results are fully consistent with, and complementary to, existing theoretical investigations of fully heavy tetraquark spectroscopy.

TABLE X. Comparison of predicted masses (in MeV) for various $cc\bar{c}\,\bar{c}$ tetraquark states. Our results are listed alongside selected theoretical predictions from the literature for the ground ($1S$), first radial excitation ($2S$), second radial excitation ($3S$), first orbital excitation ($1P$), second orbital excitation ($2P$), and first $D$-wave ($1D$) levels. The cited references represent a sample of different theoretical approaches to fully charmed tetraquark spectroscopy.

| State | $n^{2S+1}L_J$ | $J^{PC}$ | Our | Ref. [37] | Ref. [65] | Others |
|---|---|---|---|---|---|---|
| $T_{(4c)}(1S)$ | $1^1S_0$ | $0^{++}$ | 6162 | 5966 | 6190 | 6200 [1], 6477 [74] |
| | $1^3S_1$ | $1^{+-}$ | 6243 | 6042 | 6271 | 6051 [64], 6528 [74] |
| | $1^5S_2$ | $2^{++}$ | 6339 | 6194 | 6367 | 6223 [64], 6573 [74] |
| $T_{(4c)}(2S)$ | $2^1S_0$ | $0^{++}$ | 6754 | 6542 | 6782 | 6663 [75], 6642 [76] |
| | $2^3S_1$ | $1^{+-}$ | 6788 | 6577 | 6816 | 6675 [75], 6654 [76] |
| | $2^5S_2$ | $2^{++}$ | 6840 | 6647 | 6868 | 6698 [75], 6676 [76] |
| $T_{(4c)}(2S)$ | $3^1S_0$ | $0^{++}$ | 7231 | 6899 | 7269 | 7010 [76], 7296 [77] |
| | $3^3S_1$ | $1^{+-}$ | 7259 | 6917 | 7287 | 7017 [76], 7300 [77] |
| | $3^5S_2$ | $2^{++}$ | 7305 | 6953 | 7333 | 7030 [76], 7320 [77] |
| $T_{(4c)}(1P)$ | $1^1P_1$ | $1^{--}$ | 6603 | 6521 | 6631 | 6553 [76], 6577 [75] |
| | $1^3P_0$ | $0^{-+}$ | 6600 | 6476 | 6628 | 6460 [76], 6480 [75] |
| | $1^3P_1$ | $1^{-+}$ | 6606 | 6507 | 6634 | 6554 [76], 6577 [75] |
| | $1^3P_2$ | $2^{-+}$ | 6616 | 6545 | 6644 | 6587 [76], 6610 [75] |
| | $1^5P_1$ | $1^{--}$ | 6607 | 6463 | 6635 | 6459 [76], 6495 [75] |
| | $1^5P_2$ | $2^{--}$ | 6620 | 6514 | 6648 | 6577 [76], 6600 [75] |
| | $1^5P_3$ | $3^{--}$ | 6636 | 6570 | 6664 | 6623 [76], 6641 [75] |
| $T_{(4c)}(2P)$ | $2^1P_1$ | $1^{--}$ | 7063 | 6870 | 7091 | 6940 [78], 7143 (Id), 8174 (IId) [79] |
| | $2^3P_0$ | $0^{-+}$ | 7072 | 6807 | 7100 | 6953 [78], 7130 (Id), 8162 (IId) [79] |
| | $2^3P_1$ | $1^{-+}$ | 7071 | 6851 | 7099 | 7130 (Id), 8162 (IId) [79] |
| | $2^3P_2$ | $2^{-+}$ | 7070 | 6907 | 7098 | 7130 (Id), 8162 (IId) [79] |
| | $2^5P_1$ | $1^{--}$ | 7085 | 6791 | 7113 | 6943 [78], 7134 (Id), 8166 (IId) [79] |
| | $2^5P_2$ | $2^{--}$ | 7085 | 6866 | 7113 | 7134 (Id), 8166 (IId) [79] |
| | $2^5P_3$ | $3^{--}$ | 7084 | 6950 | 7112 | 7134 (Id), 8166 (IId) [79] |
| $T_{(4c)}(1D)$ | $1^1D_2$ | $2^{++}$ | 6893 | 6776 | ... | 6827 [78], 6832 [80] |
| | $1^3D_1$ | $1^{+-}$ | 6881 | 6777 | ... | 6829 [78], 6833 [80] |
| | $1^3D_2$ | $2^{+-}$ | 6892 | 6774 | ... | 6835 [80] |
| | $1^3D_3$ | $3^{+-}$ | 6904 | 6777 | ... | 6844 [80] |
| | $1^5D_0$ | $0^{++}$ | 6871 | 6784 | ... | 6827 [78], 6848 [80] |
| | $1^5D_1$ | $1^{++}$ | 6876 | 6780 | ... | 6827 [78], 6851 [80] |
| | $1^5D_2$ | $2^{++}$ | 6887 | 6775 | ... | 6827 [78], 6857 [80] |
| | $1^5D_3$ | $3^{++}$ | 6901 | 6774 | ... | 6863 [80] |
| | $1^5D_4$ | $4^{++}$ | 6917 | 6778 | ... | 6870 [80] |

TABLE XI. Comparison of predicted masses (in MeV) for various $bb\bar{b}\bar{b}$ tetraquark states. Our results are listed alongside selected theoretical predictions from the literature for the ground ($1S$), first radial excitation ($2S$), second radial excitation ($3S$), first orbital excitation ($1P$), second orbital excitation ($2P$), and first $D$-wave ($1D$) levels. The cited references represent a sample of different theoretical approaches to fully bottom tetraquark spectroscopy.

| State | $n^{2S+1}L_J$ | $J^{PC}$ | Ours | Ref. [37] | Ref. [65] | Others |
|---|---|---|---|---|---|---|
| $T_{(4c)}(1S)$ | $1^1S_0$ | $0^{++}$ | 19286 | 18719 | 19315 | 18754 [64], 18748 [78] |
| | $1^3S_1$ | $1^{+-}$ | 19292 | 18734 | 19320 | 18808 [64], 18828 [78] |
| | $1^5S_2$ | $2^{++}$ | 19302 | 18763 | 19331 | 18916 [64], 18900 [78] |
| $T_{(4c)}(2S)$ | $2^1S_0$ | $0^{++}$ | 19652 | 19065 | 19680 | 19335 [78], 19719 [80] |
| | $2^3S_1$ | $1^{+-}$ | 19654 | 19067 | 19682 | 19366 [78], 19722 [80] |
| | $2^5S_2$ | $2^{++}$ | 19659 | 19071 | 19687 | 19398 [78], 19726 [80] |
| $T_{(4c)}(3S)$ | $3^1S_0$ | $0^{++}$ | 19913 | 19188 | 19941 | 19726 [81], 19887 [77] |
| | $3^3S_1$ | $1^{+-}$ | 19915 | 19189 | 19943 | 19733 [81], 19889 [77] |
| | $3^5S_2$ | $2^{++}$ | 19919 | 19191 | 19947 | 19736 [81], 19898 [77] |
| $T_{(4c)}(1P)$ | $1^1P_1$ | $1^{--}$ | 19508 | 19361 | 19536 | 19281 [78], 19479 [82] |
| | $1^3P_0$ | $0^{-+}$ | 19505 | 19342 | 19533 | 19288 [78], 19500 [82] |
| | $1^3P_1$ | $1^{-+}$ | 19507 | 19355 | 19535 | 19496 [82] |
| | $1^3P_2$ | $2^{-+}$ | 19511 | 19371 | 19539 | 19492 [82] |
| | $1^5P_1$ | $1^{--}$ | 19506 | 19338 | 19534 | 19288 [78], 19603 [82] |
| | $1^5P_2$ | $2^{--}$ | 19510 | 19359 | 19538 | 19476 [82] |
| | $1^5P_3$ | $3^{--}$ | 19517 | 19384 | 19545 | 19617 [82] |
| $T_{(4c)}(2P)$ | $2^1P_1$ | $1^{--}$ | 19792 | 19669 | 19820 | 19597 [78] |
| | $2^3P_0$ | $0^{-+}$ | 19793 | 19646 | 19821 | 19602 [78] |
| | $2^3P_1$ | $1^{-+}$ | 19793 | 19662 | 19821 | ... |
| | $2^3P_2$ | $2^{-+}$ | 19794 | 19682 | 19822 | ... |
| | $2^5P_1$ | $1^{--}$ | 19795 | 19639 | 19823 | ... |
| | $2^5P_2$ | $2^{--}$ | 19795 | 19667 | 19823 | ... |
| | $2^5P_3$ | $3^{--}$ | 19796 | 19697 | 19824 | ... |
| $T_{(4c)}(1D)$ | $1^1D_2$ | $2^{++}$ | 19687 | 19593 | 19715 | 19510 [78], 19669 [80] |
| | $1^3D_1$ | $1^{+-}$ | 19682 | 19593 | 19710 | 19511 [78], 19671 [80] |
| | $1^3D_2$ | $2^{+-}$ | 19686 | 19592 | 19714 | 19672 [80] |
| | $1^3D_3$ | $3^{+-}$ | 19692 | 19593 | 19720 | 19675 [80] |
| | $1^5D_0$ | $0^{++}$ | 19677 | 19596 | 19705 | 19513 [78], 19677 [80] |
| | $1^5D_1$ | $1^{++}$ | 19679 | 19595 | 19707 | 19512 [78], 19678 [80] |
| | $1^5D_2$ | $2^{++}$ | 19683 | 19593 | 19711 | 19510 [78], 19680 [80] |
| | $1^5D_3$ | $3^{++}$ | 19689 | 19593 | 19717 | 19684 [80] |
| | $1^5D_4$ | $4^{++}$ | 19696 | 19594 | 19724 | 19686 [80] |

## V. CONCLUSION

In this work, we presented a systematic investigation of the mass spectra for fully heavy tetraquarks $Q_1Q_2\bar{Q}_1\bar{Q}_2$ ($Q_{1,2} \in \{b, c\}$) within the framework of a nonrelativistic diquark-antidiquark model. Our primary objective was to compute the ground and excited state masses spanning radial ($nS$ up to $4S$) and orbital ($1P$, $2P$, $3P$, $1D$) excitations and to analyze their stability against strong fall-apart decays by comparing predicted masses with the relevant meson-meson thresholds. The results are directly confronted with recent experimental data from the LHC collaborations (LHCb, CMS, and ATLAS) to provide plausible interpretations for the observed exotic resonances in the di-$J/\psi$ spectrum. We modeled the tetraquark as a two-body system composed of an axial-vector diquark $[Q_1Q_2]_{s=1}$ and an axial-vector antidiquark $[\bar{Q}_1\bar{Q}_2]_{s=1}$, interacting via a modified version of the Cornell potential that incorporates a screened confining term. This screened potential, originally successful in describing heavy quarkonia, accounts for vacuum polarization and string-breaking effects at large separations a crucial feature for radially and orbitally excited states. The time-independent Schrödinger equation was solved numerically using the three-point central difference method. Spin-dependent interactions were treated nonperturbatively, with the spin-spin term implemented via a Gaussian-smeared delta function to improve phenomenological agreement. The model parameters ($\alpha_s$, $b$, $\eta$, $\sigma$) and constituent quark masses were determined through a $\chi^2$ fit to the experimentally known masses of heavy mesons, ensuring a reliable baseline for the diquark and tetraquark mass calculations.

We obtained comprehensive mass spectra for the fully charmed $cc\bar{c}\bar{c}$, mixed heavy $bc\bar{b}\bar{c}$, and fully bottom $bb\bar{b}\bar{b}$ tetraquark systems. The spectra exhibit clear systematic trends. The ground-state ($1S$) masses are approximately 6.16–6.34 GeV for $cc\bar{c}\bar{c}$, 12.81–12.86 GeV for $bc\bar{b}\bar{c}$, and 19.29–19.30 GeV for $bb\bar{b}\bar{b}$. Radial and orbital excitations follow a regular pattern, with excitation energies on the order of several hundred MeV, consistent with the softening of the screened confining potential. Spin-dependent splittings within each multiplet are relatively small (tens of MeV), as expected for heavy-quark systems. A critical aspect of our study was the evaluation of each tetraquark state's mass relative to the lowest relevant meson-meson dissociation thresholds ($\Delta = M_{\text{tetraquark}} - M_{\text{threshold}}$). Looking at the fully charmed tetraquark $cc\bar{c}\bar{c}$, it was observed that the ground state $T_{(4c)0^{++}}(1S)$ at 6162 MeV lies 32 MeV below the di-$J/\psi$ threshold ($\Delta < 0$), suggesting a narrow state suppressed from fall-apart decay. Higher excitations ($1P$, $2S$, $1D$, $2P$, and $3S$) generally lie above threshold ($\Delta > 0$), implying broader widths due to open decay channels. For the doubly charmed bottom tetraquark $bc\bar{b}\bar{c}$, all predicted states are substantially above the lowest thresholds ($\Delta$300 MeV), indicating they are likely broad resonances. For fully bottom $bb\bar{b}\bar{b}$ tetraquarks, all states lie from 360 to 1000 MeV above the di-$\Upsilon(1S)$ threshold, predicting very broad resonances and explaining the current nonobservation in experiments. However, some high-spin $D$-wave states lie close to specific thresholds [e.g., $\Upsilon(1S)\Upsilon_2(1D)$], which could permit narrower widths. Our calculated spectra provide compelling candidate assignments for the recently observed structures in the di-$J/\psi$ mass spectrum:

- (i) $X(6200)$. Identified with the subthreshold $T_{(4c)0^{++}}(1S)$ state at 6162 MeV. Its negative $\Delta$ supports a narrow width, consistent with the ATLAS observation.
- (ii) $X(6600)$. Associated with a multiplet of $1P$-wave states ($0^{-+}$, $1^{-+}$, $2^{-+}$) around 6620–6640 MeV, where small positive $\Delta$ values align with the observed broad width.
- (iii) $X(6900)$. Interpreted as a dense set of overlapping $1D$-wave states (e.g., $2^{++}$, $3^{+-}$, $3^{++}$, $4^{++}$) in the range 6869–6918 MeV, matching the mass and width reported by LHCb and CMS.
- (iv) $X(7200)/X(7300)$. Assigned to radially excited $3S$-wave states ($0^{++}$, $1^{+-}$, $2^{++}$) between 7189–7251 MeV, consistent with the higher-mass broad structures.

Our mass predictions show broad consistency with a range of earlier theoretical approaches, including nonrelativistic quark models, relativistic diquark-antidiquark frameworks, and potential models. While absolute masses vary by 50–150 MeV depending on parameter choices and specific approximations, the overall spectral patterns-level ordering, spin splittings, and excitation gaps are remarkably similar across different studies. This agreement underscores the robustness of the diquark-antidiquark picture for fully heavy tetraquarks and validates our use of a screened potential to describe higher excitations.

In summary, we have performed a detailed, systematic calculation of the mass spectra for fully heavy tetraquarks $cc\bar{c}\bar{c}$, $bc\bar{b}\bar{c}$, and $bb\bar{b}\bar{b}$ within a nonrelativistic diquark-antidiquark model incorporating a screened confining potential. Our results successfully map the observed exotic resonances $X(6200)$, $X(6600)$, $X(6900)$, and $X(7200/7300)$ onto specific radial, orbital, and spin excitations of the fully charmed tetraquark system. The threshold analysis elucidates the width patterns: subthreshold states are candidate narrow resonances, while those above threshold are broader, in qualitative agreement with experimental observations. For the fully bottom system, the predicted masses lie significantly above the di-$\Upsilon$ threshold, suggesting broad, hard-to-observe resonances, which explains the current lack of experimental evidence. The mixed $bc\bar{b}\bar{c}$ system, though theoretically predicted, also appears prone to rapid fall-apart decays. This study reinforces the interpretation of the recently discovered di-$J/\psi$ resonances as excited states of compact fully charmed tetraquarks. It provides a coherent spectroscopic framework that can guide future experimental searches, particularly in identifying higher-spin and orbitally excited states through their decay channels. Future work should incorporate decay width calculations directly, explore the role of configuration mixing (e.g., between $2S$ and $1D$ states), and extend the model to include coupled-channel effects, which may further refine the predictions and enhance the confrontation with precision LHC data.

## ACKNOWLEDGMENTS

This project was funded by the Swiss Government Excellence Scholarship No. 2025.0419.

The authors declare that they have no known competing financial interests or personal relationships that could have appeared to influence the work reported in this paper.

## DATA AVAILABILITY

The data that support the findings of this article are not publicly available. The data are available from the authors upon reasonable request.

## APPENDIX A: EXPECTATION VALUE OF THE TENSOR OPERATOR IN A $Q_1Q_2\bar{Q}_1\bar{Q}_2$ ($Q_{1,2} \in \{c,b\}$) TETRAQUARK STATE

In this section, we compute step-by-step the expectation value of the tensor operator $\hat{\mathbf{S}}_{12}$ in a specific tetraquark state. The state is taken from Eq. (21), representing a diquark-antidiquark configuration with total angular momentum $J_T = 2$, $M_{J_T} = 2$, total spin $S_T = 2$, and orbital angular momentum $L_T = 1$, namely, the state $|J_T = 2, M_{J_T} = 2\rangle$. The calculation that we are going to present here for $|J_T = 2, M_J = 2\rangle$ with $S_T = 2$, $L_T = 1$ is one example. The same procedure applies to all other nonvanishing tensor states, with appropriate coefficients determined by $(S_T, L_T, J_T)$. Since we are interested in states with $J^{PC}$ quantum numbers including $0^{++}$, $1^{+-}$, and $2^{++}$ for $S$-waves; $0^{-+}$, $1^{-\pm}$, $2^{-\pm}$, and $3^{--}$ for $P$-waves; and $0^{++}$, $1^{+\pm}$, $2^{+\pm}$, $3^{+\pm}$, and $4^{++}$ for $D$-waves, of the fully heavy tetraquarks, the tensor interaction $\hat{\mathbf{S}}_{12}$ splits energy levels only for states with nonzero orbital angular momentum ($L_T \geq 1$) and nonzero total spin ($S_T \geq 1$). In Table XII, we classify all possible $S$-, $P$-, and $D$-wave states and indicate which require explicit calculation of $\langle\hat{\mathbf{S}}_{12}\rangle$. The tensor operator $\hat{\mathbf{S}}_{12}$ is a rank-2 spherical tensor in both spin and orbital spaces. Its expectation value vanishes if:

(1) $L_T = 0$ ($S$-wave) because the angular integral $\int Y_0^*(\theta,\phi) Y_2^m(\theta,\phi) Y_0(\theta,\phi) \sin^2\theta d\theta d\phi = 0$,

TABLE XII. Classification of tetraquark states by orbital angular momentum $L_T$, total spin $S_T$, total angular momentum $J_T$, and parity/charge-conjugation $J^{PC}$. The seventh column shows the coupled angular-spin wave function, where $C^{J_T M_{J_T}}_{L_T M_{L_T}; S_T M_{S_T}}$ are the Clebsch-Gordan coefficients, $Y^{M_{L_T}}_{L_T}$ are the spherical harmonics, and $|S_T, M_{S_T}\rangle$ are spin eigenstates. States marked "yes" require explicit calculation of the tensor expectation value $\langle\hat{\mathbf{S}}_{12}\rangle$. Those marked "no" have vanishing tensor contributions.

| Waves | $L_T$ | $S_T$ | $J_T$ | $J^{PC}$ | $\langle\hat{\mathbf{S}}_{12}\rangle$ needed? | State notation |
|---|---|---|---|---|---|---|
| $S$ | 0 | 0 | 0 | $0^{++}$ | No | $\lvert 0,0,0,M_{J_T}\rangle = Y_0^0(\theta,\phi)\lvert 0,0\rangle$ |
| | 0 | 1 | 1 | $1^{+-}$ | No | $\lvert 0,1,1,M_{J_T}\rangle = Y_0^0(\theta,\phi)\lvert 1,M_{J_T}\rangle$ |
| | 0 | 2 | 2 | $2^{++}$ | No | $\lvert 0,2,2,M_{J_T}\rangle = Y_0^0(\theta,\phi)\lvert 2,M_{J_T}\rangle$ |
| $P$ | 1 | 0 | 1 | $1^{--}$ | No | $\lvert 1,0,1,M_{J_T}\rangle = \sum_{M_{L_T}} C^{1M_{J_T}}_{1M_{L_T};00} Y_1^{M_{L_T}}(\theta,\phi)\lvert 0,0\rangle$ |
| | 1 | 1 | 0 | $0^{-+}$ | Yes | $\lvert 1,1,0,M_{J_T}\rangle = \sum_{M_{L_T}}\sum_{M_{S_T}} C^{0M_{J_T}}_{1M_{L_T};1M_{S_T}} Y_1^{M_{L_T}}(\theta,\phi)\lvert 1,M_{S_T}\rangle$ |
| | 1 | 1 | 1 | $1^{-+}$ | Yes | $\lvert 1,1,1,M_{J_T}\rangle = \sum_{M_{L_T}}\sum_{M_{S_T}} C^{1M_{J_T}}_{1M_{L_T};1M_{S_T}} Y_1^{M_{L_T}}(\theta,\phi)\lvert 1,M_{S_T}\rangle$ |
| | 1 | 1 | 2 | $2^{-+}$ | Yes | $\lvert 1,1,2,M_{J_T}\rangle = \sum_{M_{L_T}}\sum M_{S_T} C^{2M_{J_T}}_{1M_{L_T};1M_{S_T}} Y_1^{M_{L_T}}(\theta,\phi)\lvert 1,M_{S_T}\rangle$ |
| | 1 | 2 | 1 | $1^{--}$ | Yes | $\lvert 1,2,1,M_{J_T}\rangle = \sum M_{L_T}\sum_{M_{S_T}} C^{1M_{J_T}}_{1M_{L_T};2M_{S_T}} Y_1^{M_{L_T}}(\theta,\phi)\lvert 2,M_{S_T}\rangle$ |
| | 1 | 2 | 2 | $2^{--}$ | Yes | $\lvert 1,2,2,M_{J_T}\rangle = \sum_{M_{L_T}}\sum_{M_{S_T}} C^{2M_{J_T}}_{1M_{L_T};2M_{S_T}} Y_1^{M_{L_T}}(\theta,\phi)\lvert 2,M_{S_T}\rangle$ |
| | 1 | 2 | 3 | $3^{--}$ | Yes | $\lvert 1,2,3,M_{J_T}\rangle = \sum_{M_{L_T}}\sum_{M_{S_T}} C^{3M_{J_T}}_{1M_{L_T};2M_{S_T}} Y_1^{M_{L_T}}(\theta,\phi)\lvert 2,M_{S_T}\rangle$ |
| $D$ | 2 | 0 | 2 | $2^{++}$ | No | $\lvert 2,0,2,M_{J_T}\rangle = \sum_{M_{L_T}} C^{2M_{J_T}}_{2M_{L_T};00} Y_2^{M_{L_T}}(\theta,\phi)\lvert 0,0\rangle$ |
| | 2 | 1 | 1 | $1^{+-}$ | Yes | $\lvert 2,1,1,M_{J_T}\rangle = \sum_{M_{L_T}}\sum_{M_{S_T}} C^{1M_{J_T}}_{2M_{L_T};1M_{S_T}} Y_2^{M_{L_T}}(\theta,\phi)\lvert 1,M_{S_T}\rangle$ |
| | 2 | 1 | 2 | $2^{+-}$ | Yes | $\lvert 2,1,2,M_{J_T}\rangle = \sum_{M_{L_T}}\sum_{M_{S_T}} C^{2M_{J_T}}_{2M_{L_T};1M_{S_T}} Y_2^{M_{L_T}}(\theta,\phi)\lvert 1,M_{S_T}\rangle$ |
| | 2 | 1 | 3 | $3^{+-}$ | Yes | $\lvert 2,1,3,M_{J_T}\rangle = \sum_{M_{L_T}}\sum_{M_{S_T}} C^{3M_{J_T}}_{2M_{L_T};1M_{S_T}} Y_2^{M_{L_T}}(\theta,\phi)\lvert 1,M_{S_T}\rangle$ |
| | 2 | 2 | 0 | $0^{++}$ | Yes | $\lvert 2,2,0,M_{J_T}\rangle = \sum_{M_{L_T}}\sum_{M_{S_T}} C^{0M_{J_T}}_{2M_{L_T};2M_{S_T}} Y_2^{M_{L_T}}(\theta,\phi)\lvert 2,M_{S_T}\rangle$ |
| | 2 | 2 | 1 | $1^{++}$ | Yes | $\lvert 2,2,1,M_{J_T}\rangle = \sum_{M_{L_T}}\sum_{M_{S_T}} C^{1M_{J_T}}_{2M_{L_T};2M_{S_T}} Y_2^{M_{L_T}}(\theta,\phi)\lvert 2,M_{S_T}\rangle$ |
| | 2 | 2 | 2 | $2^{++}$ | Yes | $\lvert 2,2,2,M_{J_T}\rangle = \sum_{M_{L_T}}\sum_{M_{S_T}} C^{2M_{J_T}}_{2M_{L_T};2M_{S_T}} Y_2^{M_{L_T}}(\theta,\phi)\lvert 2,M_{S_T}\rangle$ |
| | 2 | 2 | 3 | $3^{++}$ | Yes | $\lvert 2,2,3,M_{J_T}\rangle = \sum_{M_{L_T}}\sum_{M_{S_T}} C^{3M_{J_T}}_{2M_{L_T};2M_{S_T}} Y_2^{M_{L_T}}(\theta,\phi)\lvert 2,M_{S_T}\rangle$ |
| | 2 | 2 | 4 | $4^{++}$ | Yes | $\lvert 2,2,4,M_{J_T}\rangle = \sum_{M_{L_T}}\sum_{M_{S_T}} C^{4M_{J_T}}_{2M_{L_T};2M_{S_T}} Y_2^{M_{L_T}}(\theta,\phi)\lvert 2,M_{S_T}\rangle$ |

(2) $S_T = 0$ because the spin tensor operator between two spin-1 diquarks vanishes for the spin singlet.

Therefore, $\langle \hat{\mathbf{S}}_{12} \rangle \neq 0$ only for states with $L_T \geq 1$ and $S_T = 1$ or 2.

We are interested in the expectation value of the tensor operator $\hat{\mathbf{S}}_{12}$ specifically for the state:

$$\begin{aligned} |\psi\rangle &= |J_T = 2, M_{J_T} = 2\rangle \\ &= \sqrt{\frac{2}{3}} |S_T = 2, M_{S_T} = 2\rangle \otimes Y_1^0(\theta,\phi) \\ &\quad - \frac{1}{\sqrt{3}} |S_T = 2, M_{S_T} = 1\rangle \otimes Y_1^1(\theta,\phi), \end{aligned} \tag{A1}$$

where

$$|S_T = 2, M_{S_T} = 2\rangle = |m_d = 1, m_{\bar{d}} = 1\rangle, \tag{A2}$$

$$|S_T = 2, M_{S_T} = 1\rangle = \frac{1}{\sqrt{2}}(|1,1;1,0\rangle + |1,0;1,1\rangle). \tag{A3}$$

We denote

$$\begin{aligned} &|A\rangle \equiv |S_T = 2, M_{S_T} = 2\rangle, \qquad |B\rangle \equiv |S_T = 2, M_{S_T} = 1\rangle, \\ &Y_0 \equiv Y_1^0(\theta,\phi), \qquad Y_1 \equiv Y_1^1(\theta,\phi), \end{aligned} \tag{A4}$$

such that the state can simply be labeled as $|\psi\rangle = \sqrt{\frac{2}{3}} Y_0 |A\rangle - \frac{1}{\sqrt{3}} Y_1 |B\rangle$. From Eqs. (32) and (33), the tensor operator is expressed as

$$\hat{\mathbf{S}}_{12} = 4[\hat{T}_0 + \hat{T}'_0 + \hat{T}_1 + \hat{T}_{-1} + \hat{T}_2 + \hat{T}_{-2}], \tag{A5}$$

where

$$\hat{T}_0 = 2\sqrt{\frac{4\pi}{5}} Y_2^0(\theta,\phi) \hat{S}_{1z}\hat{S}_{2z}, \tag{A6}$$

$$\hat{T}'_0 = -\frac{1}{2}\sqrt{\frac{4\pi}{5}} Y_2^0(\theta,\phi)(\hat{S}_{1+}\hat{S}_{2-} + \hat{S}_{1-}\hat{S}_{2+}), \tag{A7}$$

$$\hat{T}_{\pm 1} = \pm\frac{3}{2}\sqrt{\frac{8\pi}{15}} Y_2^{\mp 1}(\theta,\phi)(\hat{S}_{1z}\hat{S}_{2\pm} + \hat{S}_{1\pm}\hat{S}_{2z}), \tag{A8}$$

$$\hat{T}_{\pm 2} = 3\sqrt{\frac{2\pi}{15}} Y_2^{\mp 2}(\theta,\phi) \hat{S}_{1\pm}\hat{S}_{2\pm}. \tag{A9}$$

Here, $\hat{S}_1$ and $\hat{S}_2$ are spin-1 operators acting on the diquark and antidiquark, respectively. For the angular integrals, we define

$$\begin{aligned} I(M'_L, m, M_L) \equiv \iint & Y_1^{M'_L *}(\theta,\phi) Y_2^m(\theta,\phi) Y_1^{M_L}(\theta,\phi) \\ & \times \sin^2\theta d\theta d\phi. \end{aligned} \tag{A10}$$

Using the triple spherical harmonic integral formula:

$$I(M'_L, m, M_L) = \sqrt{\frac{45}{4\pi}} \begin{pmatrix} 1 & 2 & 1 \\ 0 & 0 & 0 \end{pmatrix} \begin{pmatrix} 1 & 2 & 1 \\ M_L & m & M'_L \end{pmatrix}. \tag{A11}$$

The first $3j$-symbol is $\begin{pmatrix} 1 & 2 & 1 \\ 0 & 0 & 0 \end{pmatrix} = -\sqrt{\frac{2}{15}}$. Thus,

$$I(M'_L, m, M_L) = -\sqrt{\frac{3}{2\pi}} \begin{pmatrix} 1 & 2 & 1 \\ M_L & m & M'_L \end{pmatrix}. \tag{A12}$$

The only nonzero $3j$-symbols needed for our state (with $M_L, M'_L \in \{0,1\}$) are

$$\begin{aligned} &\begin{pmatrix} 1 & 2 & 1 \\ 0 & 0 & 0 \end{pmatrix} = -\sqrt{\frac{2}{15}}, \qquad \begin{pmatrix} 1 & 2 & 1 \\ 0 & 1 & 1 \end{pmatrix} = \sqrt{\frac{1}{15}}, \\ &\begin{pmatrix} 1 & 2 & 1 \\ 1 & 0 & 1 \end{pmatrix} = \sqrt{\frac{2}{15}}, \qquad \begin{pmatrix} 1 & 2 & 1 \\ 1 & -1 & 0 \end{pmatrix} = \sqrt{\frac{1}{15}}. \end{aligned} \tag{A13}$$

Hence,

$$\begin{aligned} &I(0,0,0) = \sqrt{\frac{1}{5\pi}}, \qquad I(1,1,0) = -\sqrt{\frac{1}{10\pi}}, \\ &I(1,0,1) = -\sqrt{\frac{1}{5\pi}}, \qquad I(0,-1,1) = -\sqrt{\frac{1}{10\pi}}. \end{aligned} \tag{A14}$$

Terms with $m = \pm 2$ vanish because $M'_L = M_L \pm 2$ is impossible for $M_L, M'_L \in \{0,1\}$. Therefore, $\langle T_{\pm 2} \rangle = 0$. Then, we need to compute the spin matrix elements for the different spin operators including $\hat{S}_{1z}\hat{S}_{2z}$, $\hat{S}_{1+}\hat{S}_{2-} + \hat{S}_{1-}\hat{S}_{2+}$, $\hat{S}_{1z}\hat{S}_{2+} + \hat{S}_{1+}\hat{S}_{2z}$, and $\hat{S}_{1z}\hat{S}_{2-} + \hat{S}_{1-}\hat{S}_{2z}$:

$$\langle A|\hat{S}_{1z}\hat{S}_{2z}|A\rangle = 1, \quad \langle A|\hat{S}_{1+}\hat{S}_{2-} + \hat{S}_{1-}\hat{S}_{2+}|A\rangle = 0, \tag{A15}$$

$$\langle B|\hat{S}_{1z}\hat{S}_{2z}|B\rangle = 0, \quad \langle B|\hat{S}_{1+}\hat{S}_{2-} + \hat{S}_{1-}\hat{S}_{2+}|B\rangle = 2, \tag{A16}$$

$$\langle A|\hat{S}_{1z}\hat{S}_{2z}|B\rangle = 0, \quad \langle A|\hat{S}_{1+}\hat{S}_{2-} + \hat{S}_{1-}\hat{S}_{2+}|B\rangle = 0, \tag{A17}$$

$$(\hat{S}_{1z}\hat{S}_{2+} + \hat{S}_{1+}\hat{S}_{2z})|B\rangle = 2|A\rangle, \quad (\hat{S}_{1z}\hat{S}_{2-} + \hat{S}_{1-}\hat{S}_{2z})|A\rangle = 2|B\rangle. \tag{A18}$$

Thus,

$$\langle A|\hat{S}_{1z}\hat{S}_{2+} + \hat{S}_{1+}\hat{S}_{2z}|B\rangle = 2, \qquad \langle B|\hat{S}_{1z}\hat{S}_{2-} + \hat{S}_{1-}\hat{S}_{2z}|A\rangle = 2. \tag{A19}$$

For the computation of $\langle \hat{T}_0 \rangle$ and $\langle \hat{T}'_0 \rangle$, we must keep in mind that for $\langle \hat{T}_0 \rangle$, for instance, only the $|AY_0\rangle$ term contributes. Thus, we have

$$\langle AY_0|\hat{T}_0|AY_0\rangle = 2\sqrt{\frac{4\pi}{5}}\langle AY_0|Y_2^0(\theta,\phi)\hat{S}_{1z}\hat{S}_{2z}|AY_0\rangle$$
$$= \frac{2}{3}\cdot 2\sqrt{\frac{4\pi}{5}}\cdot\sqrt{\frac{1}{5\pi}}\cdot 1 = \frac{8}{15}. \tag{A20}$$

For $\langle\hat{T}'_0\rangle$, only the $|BY_1\rangle$ term contributes, and thus

$$\langle BY_1|\hat{T}'_0|BY_1\rangle$$
$$= -\frac{1}{2}\sqrt{\frac{4\pi}{5}}\langle BY_1|Y_2^0(\theta,\phi)(\hat{S}_{1+}\hat{S}_{2-}+\hat{S}_{1-}\hat{S}_{2+})|BY_1\rangle$$
$$= \frac{1}{3}\cdot\left(-\frac{1}{2}\sqrt{\frac{4\pi}{5}}\right)\cdot\left(-\sqrt{\frac{1}{5\pi}}\right)\cdot 2 = \frac{2}{15}. \tag{A21}$$

For the computation of $\langle\hat{T}_1\rangle$ and $\langle\hat{T}_{-1}\rangle$, the only nonzero cross term is $\langle AY_0|\hat{T}_1|BY_1\rangle$ and its adjoint $\langle BY_1|\hat{T}_{-1}AY_0$. For $\langle AY_0|\hat{T}_1|BY_1\rangle$, we have

$$\langle AY_0|\hat{T}_1|BY_1\rangle = \left(\frac{3}{2}\sqrt{\frac{8\pi}{15}}\right)\cdot\left(-\sqrt{\frac{1}{10\pi}}\right)\cdot 2 = -\frac{6}{5\sqrt{3}}. \tag{A22}$$

The coefficient product is $c_1^*c_2 = \sqrt{\frac{2}{3}}\cdot(-\frac{1}{\sqrt{3}}) = -\frac{\sqrt{2}}{3}$. So, the contribution is

$$c_1^*c_2\langle AY_0|\hat{T}_1|BY_1\rangle = \left(-\frac{\sqrt{2}}{3}\right)\cdot\left(-\frac{6}{5\sqrt{3}}\right) = \frac{2\sqrt{6}}{15}. \tag{A23}$$

Similarly, for $\langle BY_1|\hat{T}_{-1}|AY_0\rangle$, we get

$$c_2^*c_1\langle BY_1|\hat{T}_{-1}|AY_0\rangle = \left(-\frac{\sqrt{2}}{3}\right)\cdot\left(\frac{6}{5\sqrt{3}}\right) = -\frac{2\sqrt{6}}{15}. \tag{A24}$$

Now, taking the sum yields

$$\langle\hat{T}_1\rangle + \langle\hat{T}_{-1}\rangle = \frac{2\sqrt{6}}{15} - \frac{2\sqrt{6}}{15} = 0. \tag{A25}$$

Finally, the total expectation value $\langle\hat{\mathbf{S}}_{12}\rangle$ for the state $|J_T = 2, M_{J_T} = 2\rangle$ given in Eq. (21) is

$$\langle\mathbf{S}_{12}\rangle = 4[\langle T_0\rangle + \langle T'_0\rangle + \langle T_1\rangle + \langle T_{-1}\rangle + \langle T_2\rangle + \langle T_{-2}\rangle]$$
$$= 4\left[\frac{8}{15} + \frac{2}{15} + 0 + 0\right]$$
$$= 4\cdot\frac{2}{3} = \frac{8}{3}. \tag{A26}$$

This demonstrates the explicit calculation of the tensor operator expectation value using the spherical tensor decomposition and angular momentum algebra.

## APPENDIX B: NUMERICAL METHOD

In most existing studies, spin-dependent interactions are treated within the framework of perturbation theory. While this approach introduces perturbative corrections to hadron masses, it does not modify the corresponding wave functions. Consequently, the feedback of spin-dependent forces on the spatial structure of the bound states is neglected. To obtain a self-consistent description in which spin-dependent interactions affect both the mass spectrum and the wave functions of hadronic states, we adopt a fully nonperturbative treatment of these interactions, following Refs. [27,36]. Within this framework, the full effective potential including the nonrelativistic screened potential as well as the spin-spin, spin-orbit, and tensor terms, is incorporated directly into the Schrödinger equation and solved numerically. This allows the resulting wave functions to consistently reflect the complete dynamics of the system. To solve the resulting nonrelativistic radial Schrödinger equation, we employ the three-point central difference method. This numerical technique provides accurate approximations for first- and second-order derivatives by using the values of the wave function at three neighboring grid points, namely, the point of interest and its two immediate neighbors. Compared with forward or backward difference schemes, the central difference method offers improved numerical stability and higher accuracy, particularly when high precision is required or when the effective potential exhibits rapid variations [35]. The radial Schrödinger equation can be written in the form

$$\frac{d^2\psi(r)}{dr^2} = -2\mu\left[E_{nLSJ} - V(r) - \frac{L(L+1)}{r^2}\right]\psi(r)$$
$$= F(r)\psi(r), \tag{B1}$$

where

$$F(r) = -2\mu\left[E_{nLSJ} - V(r) - \frac{L(L+1)}{r^2}\right], \tag{B2}$$

$\mu$ denotes the reduced mass of the system, $E_{nLSJ}$ is the binding energy, and $V(r)$ represents the full potential. According to the three-point central difference method, the second derivative of the wave function is discretized on a uniform radial grid $r_k = kh$ $(k = 0, 1, 2, \ldots)$, where $h$ is the step size. The resulting recurrence relation for the radial wave function reads [35,36]:

$$\psi(r_{k+1}) = \frac{[2+\frac{5}{6}h^2F(r_k)]\psi(r_k) - [1-\frac{1}{12}h^2F(r_{k-1})]\psi(r_{k-1})}{1-\frac{1}{12}h^2F(r_{k+1})}. \tag{B3}$$

In this work, the radial Schrödinger equation is solved point by point using this recurrence relation, starting from the origin $(r = 0)$ and extending outward toward large distances ($r \to \infty$), as adopted in Refs. [27,35,36]. A crucial aspect of this procedure is the correct treatment of the wave function behavior near the origin. When $r \to 0$ and spin-orbit and tensor interactions are neglected, the regular solution behaves as

$$\psi(r \to 0) \propto r^{L+1}. \tag{B4}$$

However, when the full effective potential is included, the radial equation contains terms proportional to $1/r$, $1/r^2$, and $1/r^3$, in addition to screening effects. In the limit $r \to 0$, the dominant contribution behaves as

$$V_{\text{eff}}(r) \propto \frac{1}{r^3}, \tag{B5}$$

which renders the short-distance behavior of $\psi(r)$ ambiguous and numerically unstable and prevents a direct application of the three-point central difference method. Indeed, as $r \to 0$, the effective potential behaves like

$$V_{\text{eff}}(r) \propto \frac{1}{r^3}. \tag{B6}$$

Such a strong singularity dominates the Schrödinger equation. Thus, the standard regular solution $\psi(r) \propto r^{L+1}$ is no longer guaranteed to be valid. As a result, the wave function may diverge, oscillate uncontrollably, or depend on arbitrary boundary prescriptions near the origin. Hence, the limit $r \to 0$ of the wave function cannot be uniquely or reliably determined, making the boundary condition at the origin ambiguous. To overcome this difficulty, we assume that within a small region $r \in (0, r_c)$ the effective potential can be approximated by a constant value,

$$V(r) \simeq \frac{1}{r_c^3}. \tag{B7}$$

Under this assumption, the regular behavior $\psi(r \to 0) \propto r^{L+1}$ is restored. As a result, a cutoff distance $r_c$ is introduced, which is treated as a phenomenological parameter and determined by fitting the calculated spectrum. The initial conditions required to start the numerical integration are then chosen as

$$\psi(0) = 0, \qquad \psi(h) = h^{L+1}, \tag{B8}$$

$$F(0)\psi(0) = \lim_{r\to 0} \frac{L(L+1)}{r^2} r^{L+1} = 2\delta_{L1}. \tag{B9}$$

For a given trial binding energy $E_{nLSJ}$, the recurrence relation allows the radial wave function $\psi(r)$ to be computed sequentially from the origin to large distances. The correct eigenvalue is determined using the shooting method. Starting from an initial trial energy $E^{(0)}_{nLSJ}$, the asymptotic behavior of the numerical solution at large $r$ can be written as a linear combination of a regular and an irregular solution. The physically acceptable bound-state solution requires the coefficient of the exponentially growing term to vanish. Accordingly, at large $r$, one may write

$$\psi(r, E^{(0)}_{nLSJ}) = \xi(E^{(0)}_{nLSJ}) \exp\left(+\sqrt{2mE^{(0)}_{nLSJ}}\, r\right). \tag{B10}$$

Similarly, for another trial energy $E^{(1)}_{nLSJ}$, one has

$$\psi(r, E^{(1)}_{nLSJ}) = \xi(E^{(1)}_{nLSJ}) \exp\left(+\sqrt{2mE^{(1)}_{nLSJ}}\, r\right). \tag{B11}$$

Assuming that $\xi(E_{nLSJ})$ is an analytic function of $E_{nLSJ}$, it can be expanded as

$$\xi(E^{(1)}_{nLSJ}) = \xi(E^{(0)}_{nLSJ}) + \xi'(E^{(0)}_{nLSJ})(E^{(1)}_{nLSJ} - E^{(0)}_{nLSJ}) + \cdots. \tag{B12}$$

If $|E^{(1)}_{nLSJ} - E^{(0)}_{nLSJ}|$ is sufficiently small, higher-order terms can be neglected, yielding

$$\xi'(E^{(0)}_{nLSJ}) = \frac{\xi(E^{(1)}_{nLSJ}) - \xi(E^{(0)}_{nLSJ})}{E^{(1)}_{nLSJ} - E^{(0)}_{nLSJ}} = \frac{\psi(r, E^{(1)}_{nLSJ}) \exp -\sqrt{2mE^{(1)}_{nLSJ}}\, r - \psi(r, E^{(0)}_{nLSJ}) \exp -\sqrt{2mE^{(0)}_{nLSJ}}\, r}{E^{(1)}_{nLSJ} - E^{(0)}_{nLSJ}}. \tag{B13}$$

For the true eigenvalue $E_{nLSJ}$, the coefficient $\xi(E_{nLSJ})$ must vanish. Therefore, the improved estimate for the binding energy is obtained as

$$E_{nLSJ} = E^{(0)}_{nLSJ} - \frac{\xi(E^{(0)}_{nLSJ})}{\xi'(E^{(0)}_{nLSJ})}. \tag{B14}$$

In the numerical implementation, this procedure is applied iteratively. Letting $E^{(1)}_{nLSJ} \to E^{(0)}_{nLSJ}$ and recomputing the wave function using Eqs. (76)–(78), the iteration is repeated until convergence is achieved. The recurrence is terminated when the condition

$$|E_{nLSJ} - E^{(0)}_{nLSJ}| \leq \epsilon \tag{B15}$$

is satisfied, where $\epsilon$ denotes the desired numerical accuracy.